\documentstyle[prl,aps,epsfig]{revtex}
\begin{document}
\title{Determination of the Hurst Exponent by use of Wavelet
    Transforms}
\author{Ingve Simonsen\footnote{Ingve.Simonsen@phys.ntnu.no}
and Alex Hansen\footnote{Alex.Hansen@phys.ntnu.no}} 
\address{Institutt for fysikk, Norges Teknisk-Naturvitenskapelige
Universitet, NTNU, N--7034 Trondheim, Norway}
\author{Olav Magnar Nes\footnote{Olav-Magnar.Nes@iku.sintef.no}}
\address{IKU Petroleum Research, N--7034 Trondheim, Norway}
\date{\today}
\maketitle
\begin{abstract}
We propose a new method for (global) Hurst exponent determination based on wavelets.
Using this method, we analyze synthetic data with predefined Hurst exponents,
fracture surfaces and data from economy. The results are compared to those
obtained with Fourier spectral analysis.  When many samples are available,
the wavelet and Fourier methods are comparable in accuracy.  However, when 
one or only a few samples are available, the wavelet method outperforms
the Fourier method by a large margin. 
\end{abstract}
\section{Introduction}\label{Sec1}
It has been known for quite some time that self-affine  surfaces are
abundant in nature. They can be found in various areas of
natural science such as surface growth~\cite{Meakin,Barabasi}, 
fractured surfaces \cite{Bouchaud}, geological structures \cite{Nes},
biological systems \cite{Vicsek}. Even the mercantile community 
has reported such structures~\cite{Bak} (and references therein). 
   
Self-affine surfaces in $d$-dimensional space are described 
by a set of up to $d-1$ roughness exponents.
To know these exponents have many important physical implications.
First of all, if we know all the roughness exponents, we have full control 
of the asymptotic statistical properties of the structure.
Furthermore, for instance in fracture propagation \cite{Bouchaud},
the exponents are essential in order to determine the universality 
class of the problem. 

It is quite difficult to estimate these exponents from experimental data.
Too often statements are made which are based on rather marginal data 
analysis. It is therefore important to search for  alternative ways of
analyzing the data --- even if the new methods are not better than those 
already in existence. More tools to analyze the data, widens the possibility 
to cross-check the conclusions.  It is in this spirit that we present in   
this paper a new method for determining roughness exponents.  The strong
point of this new method is its excellent averaging properties that makes 
it possible to extract roughness exponents with high precision even when 
one or only a few samples are available.

Traditionally, the methods used to determine the self-affine exponents
are done in $(1+1)$-dimensions ($d=2$). In this case we have a single
global (self-affine) exponent and it is usually referred to as the 
{\em Hurst\/} or roughness  exponent. Over the last decade or so,
several different methods have been developed in order to
measure this exponent from experimental data 
(see Ref.\ \cite{Schmittbuhl} and references therein). 
The most popular method is
the {\em Fourier power spectrum method\/} (FPS) and we will  use this 
method as ``reference'' here. A systematic study of the quality of this   
and  other traditional methods is found in Ref.\ \cite{Schmittbuhl}. 
It should be pointed out that the method described in this paper, such
as the FPS-method, is only valid for self-affine structure described
by a global Hurst exponent.

The wavelet transform  is an integral transform developed in the
early eighties in signal analysis and is today used in different fields
ranging from quantum physics and cosmology  to data compression technology. 
Since the late eighties, wavelets have  been an active 
research field in pure and applied mathematics, 
and large theoretical progress has been made. 
The wavelet transform behaves as a mathematical microscope which 
decomposes an input signal into amplitudes which depend on position and
scale. For this purpose localized functions, called wavelets, are being used. 
By changing the scale of the wavelet, one is able at a certain location to  
focus on details at higher and higher resolution. By taking advantage of the 
central properties of self-affine functions, we derive a scaling relation 
between wavelet amplitudes at different scales, from which the Hurst exponent 
can be extracted. The method is easily generalized to higher dimensions.    
 
To our knowledge, two papers discuss wavelet based techniques in connection 
with Hurst exponent measurements \cite{Arneodo,Jones}.
In \cite{Arneodo}, the wavelet transform modulus maxima method is introduced, 
and in \cite{Jones}, wavelet packet analysis is used to extract the Hurst 
exponent. The method we present here differs from both of these.
We would also like to mention two other papers which discuss
self-affine fractals by using wavelets \cite{Struzik1,Struzik2}.
However, these two papers are mainly  concerned about the so-called  
inverse fractal problem for self-affine fractals and not 
measuring of Hurst exponents.

This paper is organized as follows. In Sec.\ \ref{Sec2}
we review the central properties of self-affine surfaces, while in
Sec.\ \ref{Sec3} the  wavelet transform is reviewed.
Sec.\ \ref{Sec4} presents the derivation of the scaling
relation for self-affine functions in the wavelet domain.
In  Sec.\ \ref{Sec5} this scaling relation is applied to
synthetic and experimental data, and  Hurst exponents are extracted.
Finally, in Sec.\ \ref{Sec6}, our conclusion is presented.

\section{Self affine surfaces}
\label{Sec2}

As stated in the introduction, 
self-affine surfaces, which are generalizations of 
Brownian motion \cite{Feder,Falconer}, have
statistical properties characterized by a set of exponents. 
Let us assume that we have a function, $h(x)$,  of one variable
only (for simplicity), i.e., a fractional Brownian
motion \cite{Feder,Falconer}. Here $x$ is the horizontal
variable, while $h$ is the vertical one. Self-affinity is defined through 
statistical invariance under the transformation 
\begin{mathletters}
    \label{eq4}%
    \begin{eqnarray}
        x    &\rightarrow&  \lambda x,  \label{eq4a}\\
        h    &\rightarrow&  \lambda^H h. \label{eq4b}
    \end{eqnarray}
\end{mathletters}
Here $H$ is the Hurst exponent. By combining such transformations, one
can construct the affine group. Thus, self-affine surfaces are 
(statistically) invariant under the affine group. An alternative way of 
expressing this invariance is by the relation
\begin{eqnarray}
    h(x)    &\simeq&  \lambda^{-H} h(\lambda x). 
    \label{eq4c}
\end{eqnarray} 
Here the symbol $\simeq$ means statistical equality. This form will prove 
useful later. The Hurst exponent, $H$, is limited to the range
$0\leq H\leq 1$. The lower limit comes from requiring the surface width 
to decrease when smaller sections of the surface is studied (the opposite
being unphysical), while the upper limit comes from assuming the surface to 
be asymptotically flat.

Eqs.\ (\ref{eq4}) and (\ref{eq4c})
express that for self-affine functions
one must rescale the horizontal and vertical direction differently in
order to have statistical invariance. Thus, self-affine surfaces are by
construction anisotropic in the horizontal and vertical direction,
except when $H=1$ (self-similarity). The Hurst exponent, $H$, expresses
the tendency for $dh=(dh(x)/dx)dx$ to change sign.
When $H=1/2$ (Brownian motion), the sign of $dh$ changes randomly, and the
corresponding surface possesses no spatial correlations.  
When $1/2 < H\le 1$, the sign tends not to change, while for
$0\le H < 1/2$, there is a tendency for the sign to change
(anti-correlation).  In both intervals there are long-range correlations 
falling off as a power law. Surfaces with  $H>1/2$ 
are said to be persistent, and those with $H<1/2$ are anti-persistent. 

\section{The wavelet transform}\label{Sec3}

Here we review some of the important properties  of wavelets, without
any attempt at being complete. Rather, our aim is to provide enough 
background for the discussion that follows. For a more complete treatment of 
wavelets, see e.g.\ Ref.\ \cite{Daub}.

In physics and mathematics there are many examples of problems 
which are more easily solved in a new set of coordinates (basis), 
where the Fourier transform being the most famous one.  
Such transforms consist in calculating the amplitudes 
for each basis function of the 
new domain. As long as a set of functions is complete, it can be used
as a root for an integral transform.
 
The wavelet transform is a relatively new (integral)
transform. What makes this transform special is that the set of basis
functions, known as wavelets,  are chosen to be well-localized 
(have compact support) both in space and frequency.
Thus, one has some kind of ``dual-localization''
of the wavelets. This contrasts the situation met 
for the Fourier Transform where one only has 
``mono-localization'', meaning that localization in both position 
and frequency simultaneously is not possible.  

The wavelets are parameterized by a {\em scale parameter\/} 
(dilation parameter) 
$a>0$, and a {\em translation parameter\/} $-\infty< b <\infty$. 
What makes the wavelet transform remarkable is that the wavelet
basis can be constructed from one single function $\psi(x)$ according to
\begin{equation}\label{eq1}%
    \psi_{a;b}(x) = \psi\left(\frac{x-b}{a}\right).
\end{equation} 
In usual terminology, $\psi(x)$ is the mother function or 
{\em analyzing wavelet.\/}

Given a function $h(x)$, the (continuous) wavelet transform is defined as
\begin{eqnarray}\label{eq2}%
    {\cal W}[h](a,b) &=& \frac{1}{\sqrt{a}} 
           \int_{-\infty}^{\infty}
           \psi_{a; b}^{*}(x)\; h(x)\; dx
\end{eqnarray}
Here $\psi^{*}(x)$ denotes the complex conjugate of $\psi(x)$.
We should emphasize that some authors use a somewhat different
definition when it comes to the prefactor.  The specific formulas we derive
further on in analyzing the self-affine surfaces depends on the definition
we have chosen.

In order for a function $\psi(x)$ to be usable as an analyzing wavelet, one 
must demand that it has
zero mean. However, in nearly all applications it is in addition required
to be orthogonal to some lower order polynomials, i.e.,
\begin{equation}\label{eq3}%
      \int_{-\infty}^{\infty} x^m\psi(x) \; dx = 0, \hspace{1cm}
             0\leq m \leq n.
\end{equation}
Here the upper limit $n$ is related to what is called the order of the 
wavelet.

Unlike for instance the more familiar Fourier transform, the wavelet
transform is not completely specified before the 
analyzing wavelet (i.e., the basis)  is given. 
There is a large number of possible candidates, but we will concentrate
exclusively on one of the most popular families, namely the Daubechies
wavelet family\ \cite{Daub}.

In order for the wavelet transform to be useful for numerical
calculations, it has to be accompanied by an effective numerical 
implementation. Such an algorithm was developed by Mallat, 
and the resulting transform is known as the discrete wavelet transform 
\cite{Press}.
 
\section{The averaged wavelet coefficient method}\label{Sec4}

As was shown in  Sec.\ \ref{Sec2}, the defining feature of self-affine 
profiles is the scaling property (cf.\ Eqs. (\ref{eq4}) and (\ref{eq4c})). 
According to Eq.\ (\ref{eq4c}) one should have
${\cal W}[h(x)](a,b) \simeq  {\cal W}[\lambda^{-H}h(\lambda x)](a,b)$ for a
self-affine function $h(x)$ in the wavelet domain. 
Here, in expressions like ${\cal W}[h(x)](a,b)$, 
we have included the $x$-dependence explicitly for convenience.

Hence after a simple change of variable one has:
\begin{eqnarray}\label{eq5}%
    {\cal W}[h(x)](a,b) 
        &\simeq&  
             {\cal W}[\lambda^{-H}h(\lambda x)](a,b) \nonumber\\
        &=& 
        \frac{1}{\sqrt{a}} 
           \int_{-\infty}^{\infty} 
               \lambda ^{-H} h(\lambda x)\; 
               \psi^{*}\left(\frac{x-b}{a}\right)
       \;dx \nonumber\\
    &=&
       \lambda^{-\frac{1}{2}-H} \;\frac{1}{\sqrt{\lambda a}} 
           \int_{-\infty}^{\infty} 
               h(x') \;\psi^{*}\left(\frac{x'-\lambda b}{\lambda a}\right)
       \;dx' \nonumber\\
    &=& \lambda^{-\frac{1}{2}-H} \; 
          {\cal W}[h(x)](\lambda a,\lambda b). \nonumber
\end{eqnarray}
Thus, we have 
\begin{eqnarray}\label{eq6}%
     {\cal W}[h](\lambda a,\lambda b)  &\simeq& 
        \lambda^{\frac{1}{2}+H} \; 
           {\cal W}[h](a,b).
\end{eqnarray}
Note that this scaling relation relies heavily on the
definition\ (\ref{eq2}), so for other definitions of the wavelet
transform,  this
equation must be changed accordingly.
What the scaling relation\ (\ref{eq6}) expresses,
is that if we perform an (isotropic)
rescaling (with factor $\lambda$) of the wavelet domain
of a self-affine function, 
this is the same as rescaling the wavelet amplitude
of the original domain with a factor $\lambda^{\frac{1}{2}+H}$.

From the definition of the wavelet transform, it follows directly that
the wavelet domain of a one-dimensional function is two-dimensional; 
one dimension corresponding to scale and the other to space (translation).
So, for instance for a specified scale, we have an infinite number of
amplitudes corresponding to various translation parameters $b$.
When one is analyzing self-affine functions, like any fractal, it is
the scale rather then the translation which is of general interest to us. 
With this in mind, we propose to {\em average out} the dependency on the
translation parameter in the wavelet domain in order to find 
a representative amplitude  at a given scale. 
We suggest to use the following formula for the average 
\begin{eqnarray}\label{eq7}%
      W[h](a)&=&
    \left< \; \left|{\cal W}[h](a,b) \right| \; \right>_b,
\end{eqnarray}
where $\left< \cdot \right>_b$ is the standard arithmetic mean 
value operator with respect to the variable $b$. Here one could have
choosen some other kind of averaging procedure such as 
geometrical or harmonic means. The absolute value is included in order
to get some kind of a ``wavelet energy''. The main point is that one gets, by
averaging the absolute value of the wavelet coefficients, a representative 
``wavelet energy'' at a given scale. 
If the dataset is missing data or containing clear anomalies
in a region, the average could still be taken over wavelet 
coefficients corresponding to wavelets localized outside 
the ``damaged'' region. However, by doing so we have
to drop some of the largest scales completely because they
inevitablely will include the unwanted region. 
The Fourier method does not have this nice property, and a missing
data region will destroy the whole dataset.

Hence the scaling relation (\ref{eq6}) becomes 
\begin{eqnarray}\label{eq7a}%
     W[h](\lambda a) 
           &\simeq& \lambda^{\frac{1}{2}+H}\;  W[h](a).
\end{eqnarray}

The strategy for the data-analysis should now be clear:
(1) Wavelet transform the data into the wavelet domain. 
(2) Calculate the averaged wavelet coefficient 
$W[h](a)$ according to Eq.\ (\ref{eq7}).
(3) Plot $W[h](a)$ against
scale $a$ in a log-log plot.  
A scaling regime consisting of a straight line in this plot implies 
a self-affine behavior of the data. The slope of this straight  line
is $\frac{1}{2}+H$.

We call this simple method
the {\em Averaged  Wavelet Coefficient} (AWC) method.

\section{Data analysis with the Averaged Wavelet Coefficient (AWC)
    method}\label{Sec5}

\subsection{Synthetic data}

We are now in position to test our scaling relation (\ref{eq7a}) on
artificial and real data, and estimate the corresponding Hurst exponents.
We start by performing an AWC-analysis on a set of
artificially generated self-affine profiles with predefined Hurst exponents.
The self-affine generator used in this work is the  Voss 
algorithm\ \cite{Voss}, which also is known as the (iterated) midpoint 
displacement algorithm.
The quality of a given analyzing method is assessed by the 
difference between the Hurst exponent chosen for  the
Voss-generator,  and the  estimated value from the analysis.

In our first illustration of the practical performance of the
AWC method, we have generated $N=100$ artificial profiles with Hurst
exponent $H=0.7$ and length $L=4096$. 
The wavelet used here and from now on if nothing different is
indicated, is the Daubechies wavelet of order 12 (Daub12).
We will later demonstrate that this choice of wavelet order, for the
Daubechies family\footnote{For wavelets of other types than the
    Daubechies family, the wavelet order $n$ has to be larger then the 
    Hoelder exponent of the strongest singularity present \cite{Mallat,Muzy}.}
For each sample the mean drift of the profile,
defined as the line connecting its first and last point, is subtracted. 
In Fig.\ \ref{fig1} the results are
presented for both the AWC- and Fourier power spectrum (FPS) density method. 
In both cases a
straight line fit is performed to the (log-log) data, with resulting slopes 
of respectively $1.19\pm0.01$ and $-2.39\pm0.02$. 
Theoretically these slopes should be $\frac{1}{2}+H=1.2$
(cf. Eq.\ (\ref{eq7a})) and $-(1+2H)=-2.4$ \ \cite{Feder,Falconer} 
for respectively the AWC- and FPS-method. 
Hence the corresponding estimated Hurst exponents,
in obvious notation, become $H_{W}=0.69\pm0.01$ and $H_F=0.70\pm0.01$.  
Here we should emphasize that the
errors indicated, are only the regression errors in the actual
region. Errors due to different choices of regression regions 
are not included even if they typically are larger than the regression
error itself. The quality of the fit is indicated
in Figs.\ \ref{fig1}a and\ \ref{fig1}c, by including the fitting function. 
Empirically we would expect the {\em total\/} error of the Fourier power 
spectral density method to be larger then the corresponding
error of the AWC method. This is so because the linear scaling 
region is smallest for the FPS-method.
In order to quantify the behavior for different Hurst exponents, we have
performed a corresponding analysis to the above, for various  
exponents in the range $0<H<1$. 
The results are shown in Figs.\ \ref{fig1}b and \ref{fig1}d.
We observe that there are good  agreement between the actual and  estimated
exponents in the whole range of Hurst exponents independent of  method.

It is often the case that one does not have many data samples 
available for analysis, especially when dealing with experimental data.
To discuss this situation we have performed the same analysis as above,
but now with a smaller number of samples; $N=50$ and $N=5$ (Fig.\ \ref{fig2})
and $N=1$ (Fig.\ \ref{fig3}).  
Still the correspondence with the input value is relatively good. 
However, for small number of samples the uncertainty in the slope
determination becomes large, as illustrated in Figs.\ \ref{fig3}a and
\ref{fig3}c. This tendency is much more explicit for the FPS-method than for 
the AWC-approach.

One could now ask how these results depend on the specific  order 
chosen for actual wavelet. 
In Fig.\ \ref{fig4} we have included a graph showing the
AWC-function, $W[h](a)$, for different orders
(i.e. smoothness) of the Daubechies  wavelet family.
With the above comment made on the true error in the Hurst exponent
measurements, we conclude that within the actual errors the AWC method 
does not seem to be sensitive to the order of the wavelet, 
at leaste not for the Daubechies wavelet family. 
A non-systematic study with other kind of wavelets does not change
this conclusion. 

We have also investigated the situation where the length of the profiles 
varies. Our findings  are compiled in Table\ \ref{tab1}.
The results are generally in agreement with the input
value for system sizes larger or equal to $L=256$. It should also be
noted that the regression error (not necessarily the actual error) 
generally decreases with increasing system size.
This is as expected because when the system size increases the scaling
region becomes larger, resulting in a better regression fit.

In summary, for the study of (clean) synthetic self-affine data,
we may conclude that the AWC method works  well.
It is in particular a good choice when
only a few number of samples are available, which is often the case in
experimental situations.   

\subsection{Stability against noise, drift and distortions}

All real measurements are subject to noise, and distortions.
These may have their origin in measuring uncertainties, instrumental noise 
and non-linearities that might transform the signal in some way.  The
non-linearities may come from the response of the measuring devices used.
However, it may also be that the variable studied is not the ``good'' one.
In order for a method of analyzis to be useful for real world data, it has 
to be stable with respect both to noise and distortions.

We start by studying distortions.  Suppose that rather than measuring
the {\em generic\/} self-affine function $h(x)$, we observe 
$F[h(x)]$, i.e.,     
\begin{eqnarray}
    \label{eq8}
    h(x) &\rightarrow&  F[h(x)]\;,
\end{eqnarray}
which is a one-to-one transformation.
This may for example result from distortion of the original signal 
through the instrumentation.   
Note that we have not allowed for an explicit $x$-dependence in $F$,
because this may destroy the self-affinity (more about this later). 
By other methods it can be demonstrated that the transformation (\ref{eq8}) 
does not change the Hurst exponent \cite{Nes}. 
A qualitative way to understand this result is that, since the Hurst
exponent $H$ is related to the tendency of $dh$ to change sign (see
earlier discussion), the Hurst exponent should be insensitive to
transformations of the type\ (\ref{eq8}) as long as $F$ is a relatively
smooth functional.
To demonstrate the stability of the AWC method to such distortions, we 
have performed a numerical experiment, 
where we have put $F[h(x)]=\log_{10}(h(x))$, and then
calculated the Hurst exponent from $F[h(x)]$. The result is shown in
Fig.\ \ref{fig4a}, and as can be seen, the Hurst exponent is unchanged
within the numerical errors. 
The logarithmic function is a highly non-linear function, something which
thus changes the input data dramatical, thus providing a good testing
ground of this assumption. 

In many situations one has  signals possessing some kind of drift. It has
earlier been shown that such drifts can dramatically influence the
reliability of the measured Hurst exponent \cite{Hansen}. 
In order to test our method in this respect, we have perfomed an 
analysis where we have added linear (Figs.\ \ref{drift}a and b) 
and quadratic drift  (Figs.\ \ref{drift}c and d) to the 
self-affine component of the data.
In this part of the analysis we have not subtract the line 
connecting the first and last point of our dataset 
ahead of the wavelet transform as described earlier.
For the linear case, there is only a weak dependence, independent of
scale, on the drift of the data (Figs.\ \ref{drift}a and b).  
However, for the quadratic case, the situation is somewhat different.
Here the drift creates a nice crossover between the self-affine region,
dominating at small scales, and the drift at large scales.
This can be easily see from Fig.\ \ref{drift}d. The Hurst exponent
obtained from the small-scale region is $H=0.71\pm0.05$, which
fits quite well to the exponent, $H=0.70$, of the self-affine component     
of the data. In both cases the amplitude of the drift seems to
be of secondary importance. Based on the above, we conclude that our
method seems to work quite well for data with drift.     
 
It is easy to see that if the functional $F$ has some explicit
dependence on the ``horizontal'' coordinate (i.e. $x$ in our case), 
the self-affine correlation property may be destroyed. 
Spatial noise of any kind, 
has exactly this property. Since such noise is usually always found in real 
data, it is important to investigate how sensitive the AWC-algorithm is 
to such artefacts. In order to simulate this situation, we apply the 
functional $F[h(x),x]=h(x)+\eta(x)$, where $\eta(x)$ is a noise term,  
to clean self-affine data $h(x)$, and then proceed with the AWC anaysis.
The amount of noise added to the data, i.e., the noise rate
$\chi$, is defined as 
$\chi=(\max{|\eta(x)|}-\min{|\eta(x)|})/( \max{|h(x)|}-\min{|h(x)|})$.
 
In this paper we have chosen to work with white, pink ($1/f$)
and brown ($1/f^2$) noise.
Quite recently Aguilar and coworkers have pointed out that 
scanning tunneling microscopy instrument noise is pink \cite{Aguilar}.
The quantitative effect of adding $\chi=10\% $ noise to a given self-affine 
profile is demonstrated in Fig.\ \ref{noise}.
The results of the AWC-analysis for the case $\chi=10\% $ of added
white, pink or brown noise are shown in respectively 
Fig.\ \ref{fig5.white}, Fig.\ \ref{fig5.pink} and Fig.\ \ref{fig5.brown}. 
In all cases (with $\chi =10\% $) we see that the
AWC method extracts the actual Hurst exponent quite well.
Notice that for the white noise case, only the lower scales seem to be
considerably affected, if any,  by the noise. 
It should be observed  that for $H=0.7$ (see Fig.\ \ref{fig5.white}a)
a nice crossover to the (now smaller) scaling regime is shown, while
for  $H=0.2$ (Fig.\ \ref{fig5.white}b) this crossover is not visible.
This behaviour we have found to be quite systematic in the sense
that the higher Hurst exponents (in the range $0<H<1$) the more
pronounced was the crossover, and the smaller was the (self-affine)
scaling regime. The explanation for this behaviour is the following: 
For Hurst exponents in the lower range $0<H<0.5$,
(i.e. anti-correlation) the profiles are quite spiky with 
sharp  tops and deep valleys. 
This means that the wavelet coefficients at low scales become large 
for low Hurst exponents with the consequence that the contribution 
of the noise is suppressed. 
As the Hurst exponent gets larger, and thus the 
profiles become more smooth, the effect of the noise at small scales
will becomes more and more important resulting in a well-defined
crossover. This crossover is easily seen in Fig.\ \ref{fig5.white}a.    
 
For the white noise case, we just saw that mainly the small scales were
affected, if any, by the noise. This situation is somewhat different for the
pink and brown noise cases (see Figs.\ \ref{fig5.pink} and \ref{fig5.brown}).
For these two noise types the whole region is affected without,
for $\chi =10\% $, introducing significant changes over the non-noisy
results (Figs.\ \ref{fig5.pink}--\ref{fig5.brown}b).
As the noise level increases (see Figs.\
\ref{fig5.pink}--\ref{fig5.brown}d) larger deviation from the
none-noisy case starts to emerge.
By noting that pink and brown noise is nothing but  
self-affine signals with Hurst exponents of respectively $H=0$ and $H=0.5$, 
we would expect that the estimated exponents are shifted
towards these values. This is  supported by the observation from 
Figs.\ \ref{fig5.pink}d and \ref{fig5.brown}d that the slopes, for a
noisy $H=0.7$ profiles  seems to decrease with increasing noise level $\chi$.  
It should also be observed  that there seems to be a more well-developed
crossover for the pink than the brown case. This stems from the fact
that Figs.\ \ref{fig5.pink}d and \ref{fig5.brown}d are shown for $H=0.7$,
giving the  better ``contrast'' for the pink limiting case of $H=0$.

Self-affine scaling behavior is usually only found over a limited
region of space (or time), and it is important to be able to estimate these
crossover scales. To be able to investigate the potential of the
AWC method in this respect, we have generated some artificial
self-affine profiles with $H=0.7$ and length $L=4096$ and 
used a standard 5-point
filter to destroy the self-affine correlations at
small distances (i.e. destroy correlations between 11 subsequent points). 
For the AWC method we should expect to see a crossover at scales
$a_c=\simeq 0.003$, while for the FPS-method the crossover
frequency is expected at $f_c=0.09$.   
As can be seen from Fig.\ \ref{fig5a}, this is indeed what we find.
For the largest number of sample ($N=100$) the AWC and FPS method are
equivalent, but for only one sample the crossover is most easily seen
for the wavelet method as shown in  Figs.\ \ref{fig5a}c and d.

\subsection{Real data}

As mentioned in the introduction, self-affine surfaces can be found
many places in the sciences. Here we will in particular 
discuss two quite different examples, clearly demonstrating the general 
presence of self-affine structures. 

Our first example is taken from geology, and concerns the
structure of a fractured granite surface\ \cite{Schmittbuhl_2}.  
The surface contains $2050\times 211$ data points.
One representative profile of this surface is given in
Fig.\ \ref{fig6}a. The results of the wavelet and Fourier analysis,
using the methods described earlier in this paper, are collected in
Figs.\ \ref{fig6}b and c . We see that there are nice scaling regions in
both cases indicating that the fractured granite surface is indeed 
self-affine. The Hurst exponents, obtained by a regression fit to the
scaling region, are $H_W=0.81\pm0.02$ and $H_F=0.79\pm0.03$
respectively for the AWC  and FPS method. Note that
also here only the regression error  is indicated and that
the ``true error'' is somewhat larger. The results  for the two methods are
consistent and coincide with the results of other 
studies of fracture granite surfaces \cite{Schmittbuhl_3}.
The reason for the good agreement
between the two methods, and the good quality of both  regression fits
are due mainly to the large number of samples available, as the
dataset contains 211 one-dimensional profiles of length 2050.
It is interesting to observe that the exponent for the fractured
granite surface reported here, is in complete agreement with earlier
speculations that fracture surfaces should have a universal Hurst
exponent of $H=0.8$\ \cite{Bouchaud,Alex,Lapasset} 

Unfortunately, the availability of  large number of samples is rare. 
nor is it appropriate to talk about
several samples at all in some situations. In order to illustrate this point,
we have included share prices for the Italian automobile manufacturer FIAT 
taken from the  Milan Stock Exchange for the period from the 
1st of September 1988 to the 28th of May 1991 
with three quotes per day \cite{Vidakovic} (Fig.\ \ref{fig7}a). 
Observe that here only one sample for a given time  period is available. 
The result of the corresponding analysis is given in
Figs.\ \ref{fig7}b and c. The estimated Hurst exponents are
$H_W=0.65\pm 0.03$ and $H_F=0.62\pm 0.06$ from respectively 
the AWC and FPS method.
These two results are consistent, but there are
noticeable difference in the accuracy of the regression fits for the two
methods. This is most easily seen by comparing Figs.\ \ref{fig7}b and c
by visual inspection.  The error bars associated with the FPS analysis, based 
solely on the regression analysis, are in particular underestimated.
It is difficult to identify a scaling region at all.
We find it very interesting to observe that a Hurst exponent 
$H=0.65$ is observed in the stock marked simulations by Bak et
al.\ \cite{Bak} when using the Urn model with volatility feedback.     
 
This example, is in our view, a very good example
of the power of the AWC method in cases where few samples are available, 
and we believe it to have  potential of becoming a useful method in 
practical situations. 

\section{Conclusions}\label{Sec6}

We have introduced, derived and tested a new simple method for Hurst exponent
measurements based on the wavelet transform. It has been compared to
the Fourier power spectrum method where appropriate.
We find that the two methods performs approximately equally
for large number of samples. However, for small numbers of samples this
new method outperforms the more traditional Fourier transform based
method. The AWC method are also demonstrated to handle noisy and
experimental data in a satisfactory manner.

\acknowledgements
We are grateful to J.\ Schmittbuhl for permission to use his
measurements for the fractured granite surface. We also thank B.\ Vidakovic 
for providing us with the Fiat share price data.
One of the authors (I.\ S.) thanks the Research  Council of Norway and Norsk 
Hydro AS  for financial support. This work has received support from the 
Research  Council of Norway (Program for Supercomputing) through a grant of 
computing time.

\begin{table}[htb]
    \begin{center}
        \leavevmode
        \begin{tabular}[]{rcc}
           $L$  &  $H_F$  &  $H_W$  \\
            \hline
          $  64   $ &  $ 0.57 \pm 0.04  $  &  $ 0.67\pm    0.05 $  \\
          $  128  $ &  $ 0.65 \pm 0.03  $  &  $ 0.66\pm    0.03 $  \\
          $  256  $ &  $ 0.66 \pm 0.01  $  &  $ 0.69\pm    0.02 $  \\
          $  512  $ &  $ 0.66 \pm 0.02  $  &  $ 0.70\pm    0.02 $  \\
          $  1024 $ &  $ 0.70 \pm 0.01  $  &  $ 0.70\pm    0.01 $  \\
          $  2048 $ &  $ 0.70 \pm 0.01  $  &  $ 0.70\pm    0.01 $  \\
          $  4096 $ &  $ 0.70 \pm 0.01  $  &  $ 0.70\pm    0.01 $  \\                \end{tabular}
        \caption{Estimated Hurst exponents for the FPS- ($H_F$)
            and AWC-method ($H_W$) for different system sizes, $L$.
            The predefined Hurst exponent was $H=0.7$, and   
            in all calculations the number of samples used was $N=100$. 
            The wavelet used was Daub12.}
        \label{tab1}
    \end{center}
\end{table}
\begin{figure}
    \caption{Hurst exponents estimated by (a and b) the averaged wavelet
            method ($H_W$) and (c and d) the Fourier power spectrum density 
            method ($H_F$). All errorbar in this, and later figures,
            are regression errors only. 
            (a) The AWC-function $W[h](a)$ vs scale $a$
            for Hurst exponent $H=0.7$. The solid line is the
            regression line to the scaling region. The estimated Hurst
            exponent is $H_W=0.69\pm 0.01$. (b) Wavelet estimated Hurst 
            exponents ($H_W$) for various actual Hurst exponents $H$.
            (c) The power spectrum, $P(f)$ vs frequency $f$ for 
            actual Hurst exponent  $H=0.7$. The solid line is the
            regression fit. The estimated Hurst exponent
            is $H_F=0.70\pm 0.01$. (d) Fourier estimated Hurst exponents 
            ($H_F$) for various actual Hurst exponents $H$. 
            The number of samples per data point was $N=100$, and the
            length of the profiles were $L=4096$. The same
            profiles were used for both the wavelet and Fourier analysis.}
    \label{fig1}
\end{figure}
\begin{figure}
    \caption{The same as Fig.\ 1b and d, but now
            with the following number of samples $N=50$
            (a and b) and $N=5$ (c and d).}
    \label{fig2}
\end{figure}
\begin{figure}
    \caption{The same as Fig.\ 1, but now with only
            one sample, $N=1$.}
    \label{fig3}
\end{figure}
\begin{figure}
        \caption{The AWC-function $W[h](a)$ vs scale $a$,
            for various choices of wavelet order (Daubechies family)
            as indicated in the figure. The data are 
            for self-affine profiles with Hurst exponent $H=0.7$, and 
            the number of samples used was $N=100$. The length of the
            profiles was $L=4096$. The extracted Hurst exponents were
            $H=0.68\pm0.01$ for daub4, and $H=0.70\pm0.01$ in all
            other cases. }
        \label{fig4}
\end{figure}
\begin{figure}
    \caption{(a) The AWC-function $W[g](a)$ vs scale $a$ where 
        $g(x)=log_{10}(h(x))$  and $h(x)$ is a  self-affine
        function with $H=0.7$. The number of samples used was
        $N=100$, and length of the profiles was $L=4096$.
        The solid line is the regression fit to the data, and the
        estimated Hurst exponent was $H_W=0.70\pm0.01$
        (b)  Wavelet measured  Hurst exponents, $H_W$, for various
        actual exponents in the range $0.1\leq H \leq 0.9$.}
    \label{fig4a}
\end{figure}

\begin{figure}
    \caption{The Figures show the effect of added linear (a and b) and
        quadratic (c and d) drift to the data.
        In all cases the self-affine component of the data had $H=0.7$.
        (a) One sample with linear drift ($y(x)=0.5 x$).
        (b) The AWC-function $W[h](a)$ vs scale $a$ for data
        with linear drift of the type showed in Fig.\ a. The number of samples
        used was $N=100$ and the length of the all profiles was $L=4096$.
        The full line is the regression line.
        The extracted Hurst exponent is $H=0.70\pm0.01$.
        (c and d) The same as Figs.\ (a) and (b) respectively, but now for
        quadratic drift ($y(x)=0.05 x^2$). In Fig.\ (c) we have also
        included the drift sepreately (the dashed line).
        Notice the well-developed crossover between the larger 
        and smaller scales in Fig.\ (d). 
        The extracted Hurst exponent is $H=0.71\pm0.05$. The slope,
        $\alpha$  of the regression line (dot-dashed) for the large 
        scales is $\alpha=2.29\pm0.04$.}
    \label{drift}
\end{figure}

\begin{figure}
    \caption{(a) shows a self-affine profile with $H=0.7$. To this
        profile we add $\chi=10\% $ white (b), pink (c), and brown (d)
        noise. The lower curve in Figs. (b)--(d) is the added noise.}        
    \label{noise}
\end{figure}

\begin{figure}
    \caption{The Figures show the effect of white noise added to the
            self-affine component of the data.
            The AWC-function $W[h](a)$ vs scale $a$ for the
            self-affine component of the data with $H=0.7$ and
            $H=0.2$ and a noise level $\chi = 10\%$  (a and b). 
            The estimated Hurst exponents for the curves
            shown in Figs.\ (a) and (b) were respectively
            $H_W=0.71\pm0.02$ and $H_W=0.21\pm0.01$.
            The extracted Hurst exponents with regression errors 
            for Hurst exponents in the range $0<H<1$ and step $0.1$ (c).
            The effect of the noise on the AWC-function, $W[h](a)$,
            for various noise levels $\chi$ as indicated in the figure
            (d). The Hurst exponent og the self-affine component was
            $H=0.7$ in this case. The number of samples per data point
            was  $N=100$, and the length of the profiles was $L=4096$.}
    \label{fig5.white}
\end{figure}

\begin{figure}
    \caption{The same as Fig.\ \protect\ref{fig5.white}, 
             but now pink noise added.
             The estimated Hurst exponents for the curves
             shown in Figs.\ (a) and (b) are respectively
             $H_W=0.69\pm0.01$ and $H_W=0.21\pm0.01$.} 
    \label{fig5.pink}
\end{figure}

\begin{figure}
    \caption{The same as Fig.\ \protect\ref{fig5.white},
             but now brown noise added.
             The estimated Hurst exponents for the curves
             shown in Figs.\ (a) and (b) are respectively
             $H_W=0.69\pm0.01$ and $H_W=0.19\pm0.01$.} 
    \label{fig5.brown}
\end{figure}

\begin{figure}
    \caption{The wavelet (Figs.\ (a) and (c)) and  Fourier (Figs.\ (b) and (d))
        analysis of synthetic data with Hurst exponent $H=0.7$ after
        applying a 5-point filter to them.
        In Figs.\ (a) and (b), the results are averaged over 
        $N=100$ samples, while in Figs.\ (c) and (d) only one sample
        ($N=1$) is  used. In all cases $L=4096$.
        The solid lines are the regression fits to the scaling regions.
        The crossovers are clearly seen in all cases.
        Theoretically the crossover values are $a_c=0.003$ and
        $f_c=0.09$ for the wavelet and Fourier method respectively.}
    \label{fig5a}
\end{figure}
\begin{figure}
    \caption{(a) One single representative profile from the granite
        fracture. The number of points in the profile is $L=2050$.
        (b) AWC analysis  of the  entire set of ($2050\times211$) data
        points. 
        The solid line is the regression fit to the
        scaling region. The corresponding Hurst exponent is $H_W=0.81\pm0.02$.
        (c) FPS analysis of the data. Here the solid line corresponds to
        a Hurst exponent $H_F=0.79\pm0.03$.}
    \label{fig6}
\end{figure}
\begin{figure}
    \caption{(a) Fiat share prices taken from the Milan Stock Exchange
            for the period from the 1st of September 1988 (day 1) to
            the 28th of May 1991, with three observation  per day.
            (b) The result of the wavelet analysis for the data in (a).
            The estimated Hurst exponent, corresponding to the solid
            line, is $H_W=0.65\pm 0.03$.
            (c) The result of the Fourier analysis for the data in (a).
            The Hurst exponent in this case  is $H_F=0.62\pm 0.06$.
            Note the more well-behaved scaling region for the
            wavelet method  as compared to the Fourier method.}
    \label{fig7}
\end{figure}


\newpage
\setcounter{figure}{1}
\newcommand{\mycaption}[2]{\begin{center}{\bf Figure \thefigure}\\{#1}\\{\em #2}\end{center}\addtocounter{figure}{1}}
\newcommand{\myauthor}{Simonsen, Hansen and Nes}
\newcommand{\mytitle}{Determination of the Hurst Exponent by use of Wavelet
    Transforms}


\begin{figure}
    \begin{center}
        \begin{tabular}{@{}c@{\hspace{1.0cm}}c@{}}
            \epsfig{file=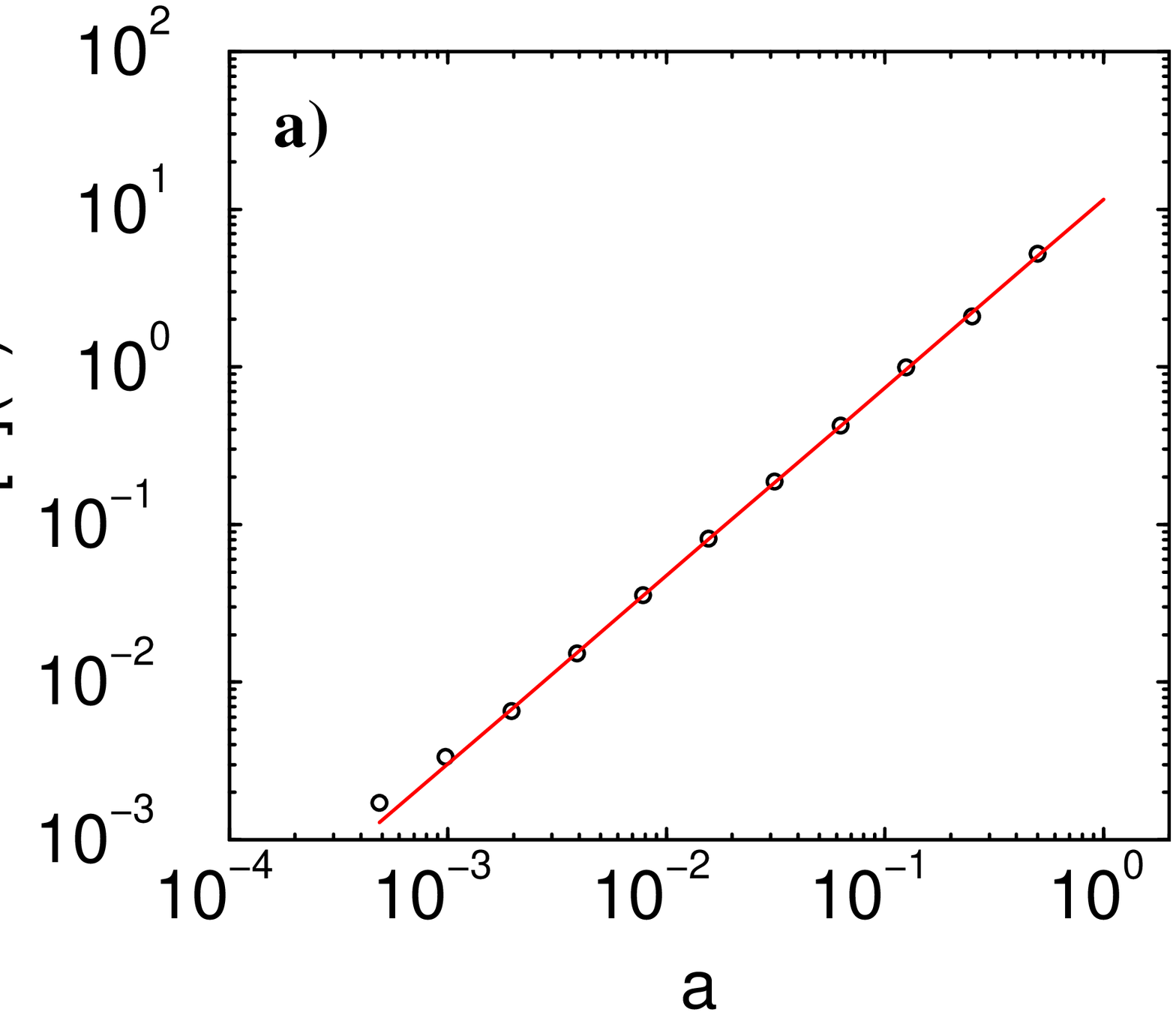,width=8.5cm,height=8.5cm} &
            \epsfig{file=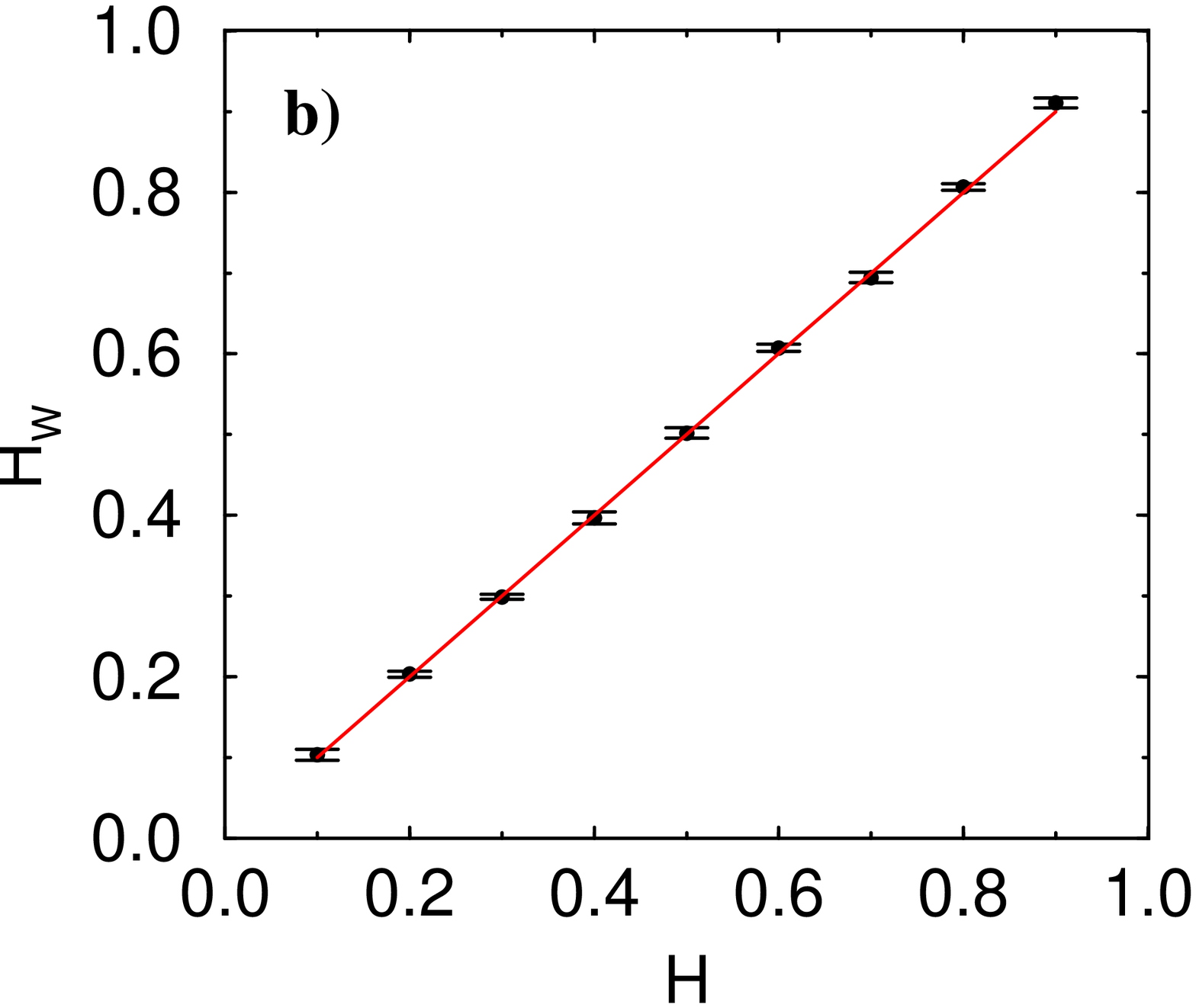,width=8.5cm,height=8.5cm}\\
            \epsfig{file=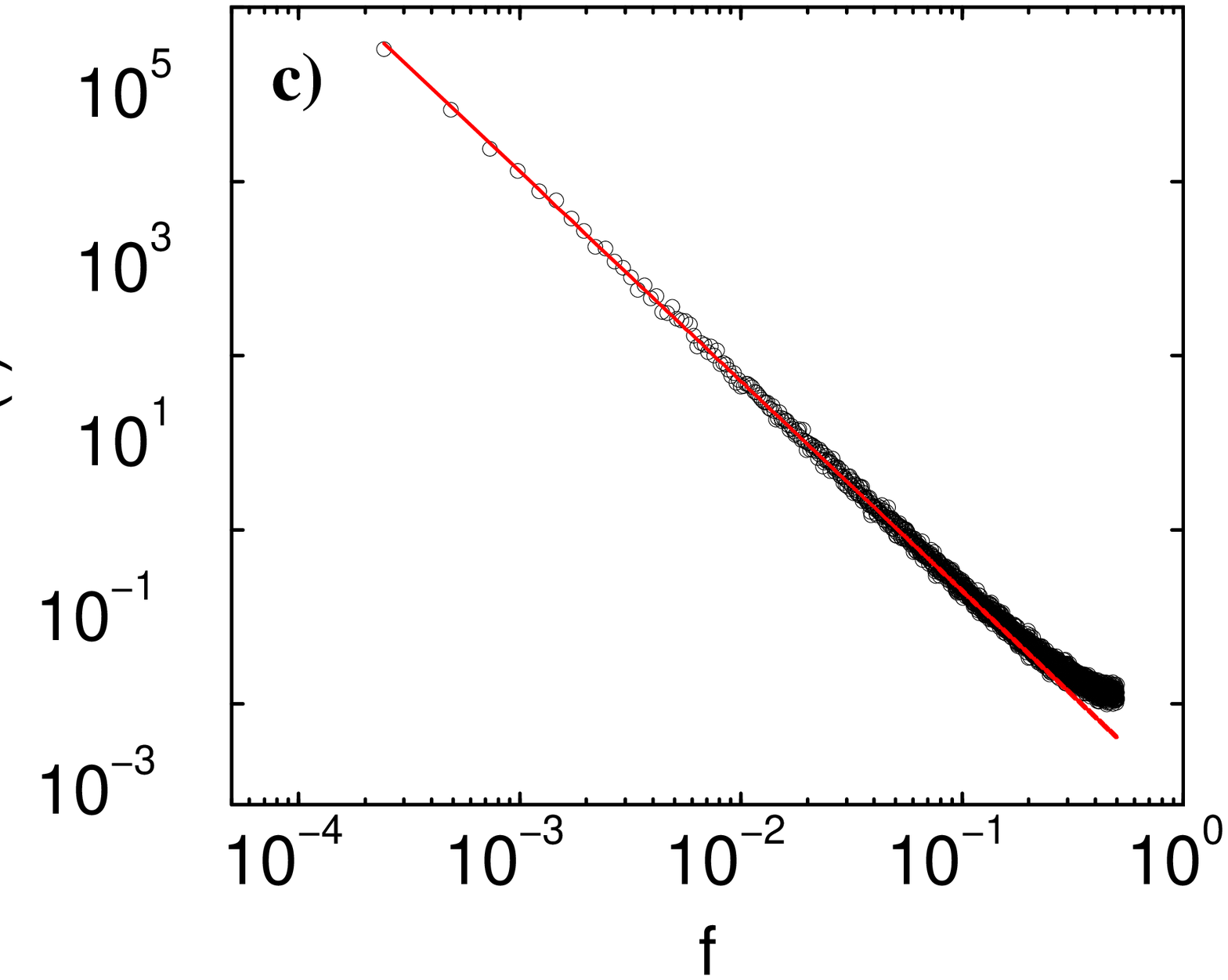,width=8.5cm,height=8.5cm} &
            \epsfig{file=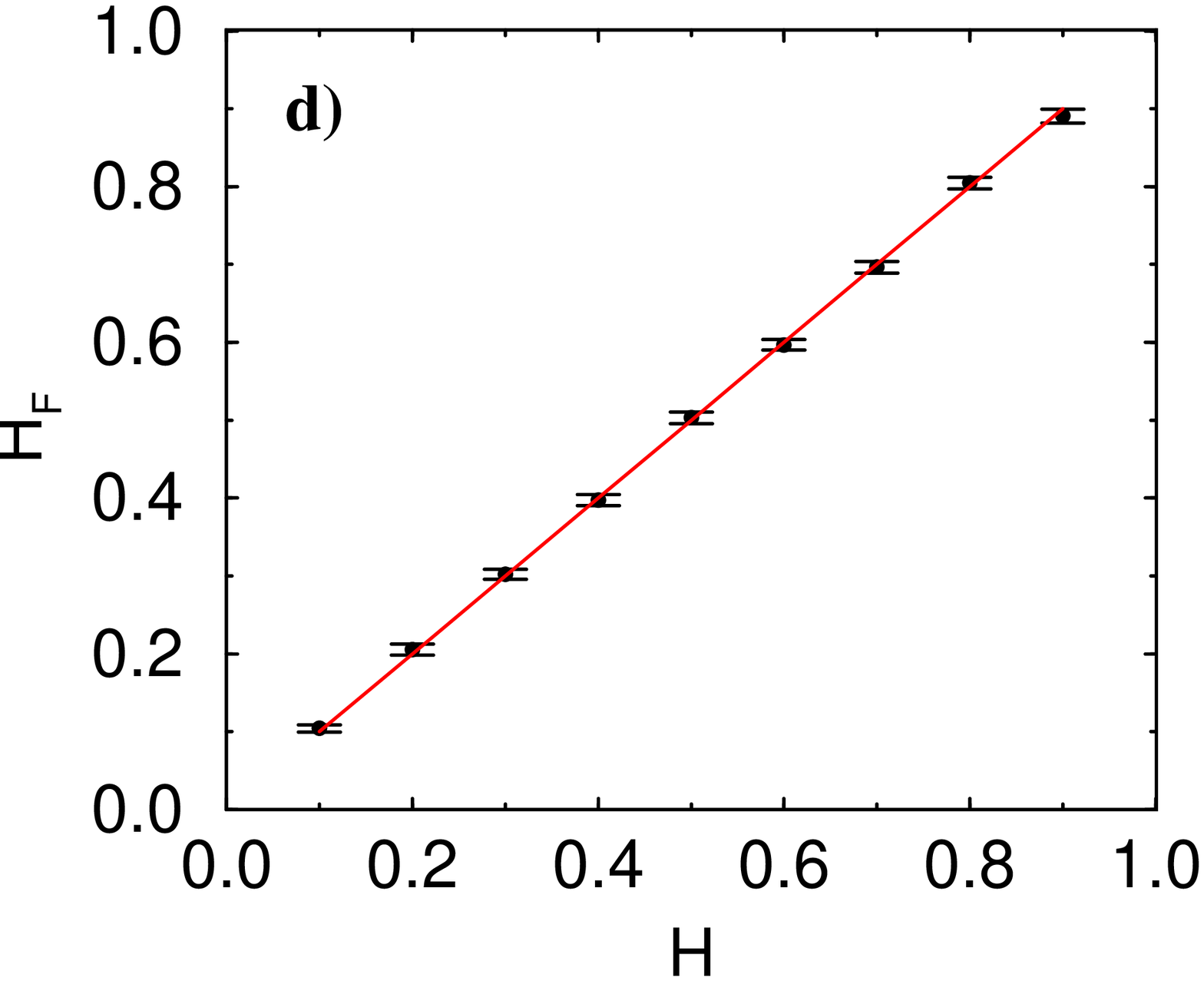,width=8.5cm,height=8.5cm} 
        \end{tabular}
    \end{center}
    \mycaption{\myauthor}{\mytitle}
\end{figure}

\begin{figure}
    \begin{center}
        \begin{tabular}{@{}c@{\hspace{1.0cm}}c@{}}
            \epsfig{file=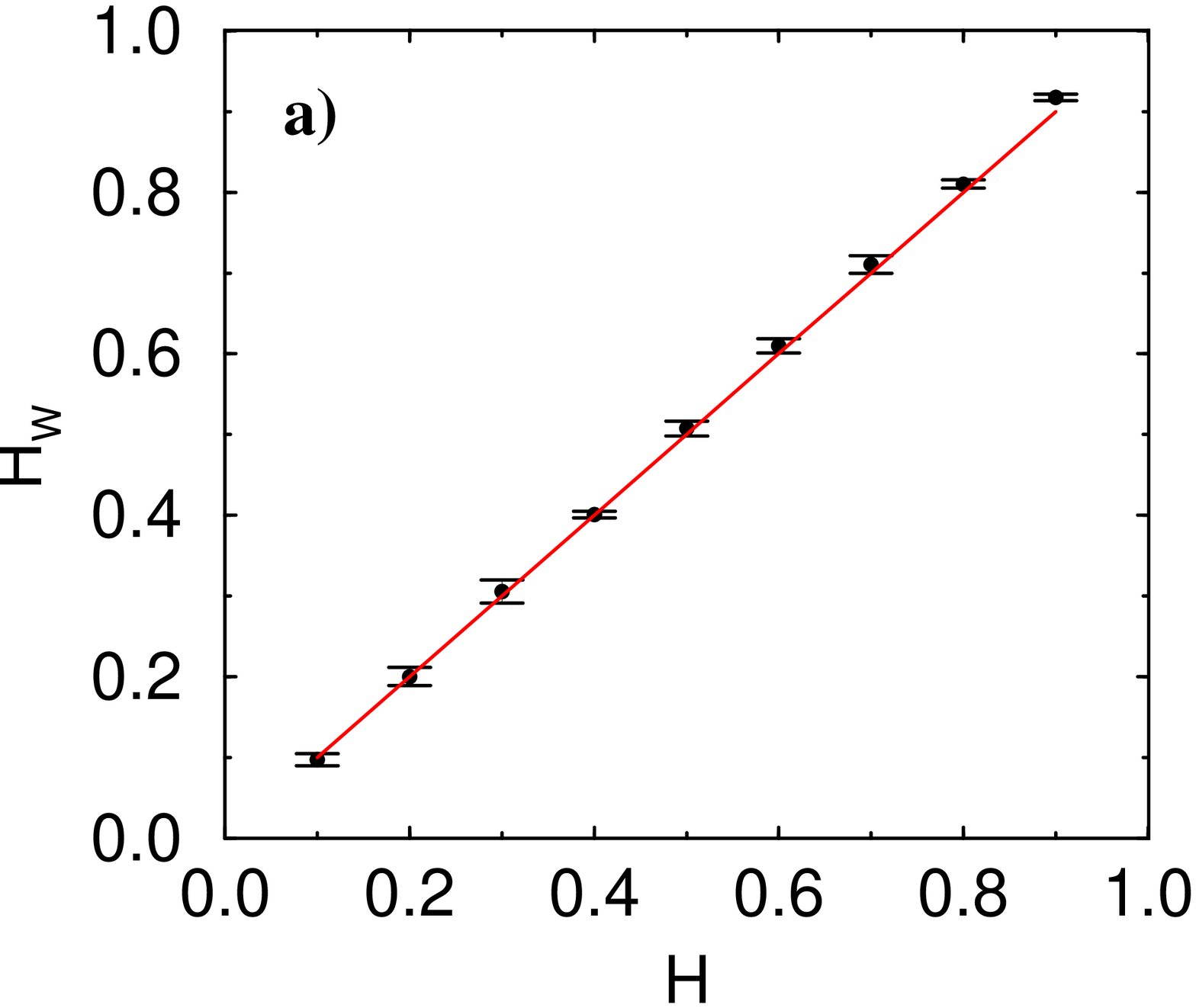,width=8.5cm,height=8.5cm} &
            \epsfig{file=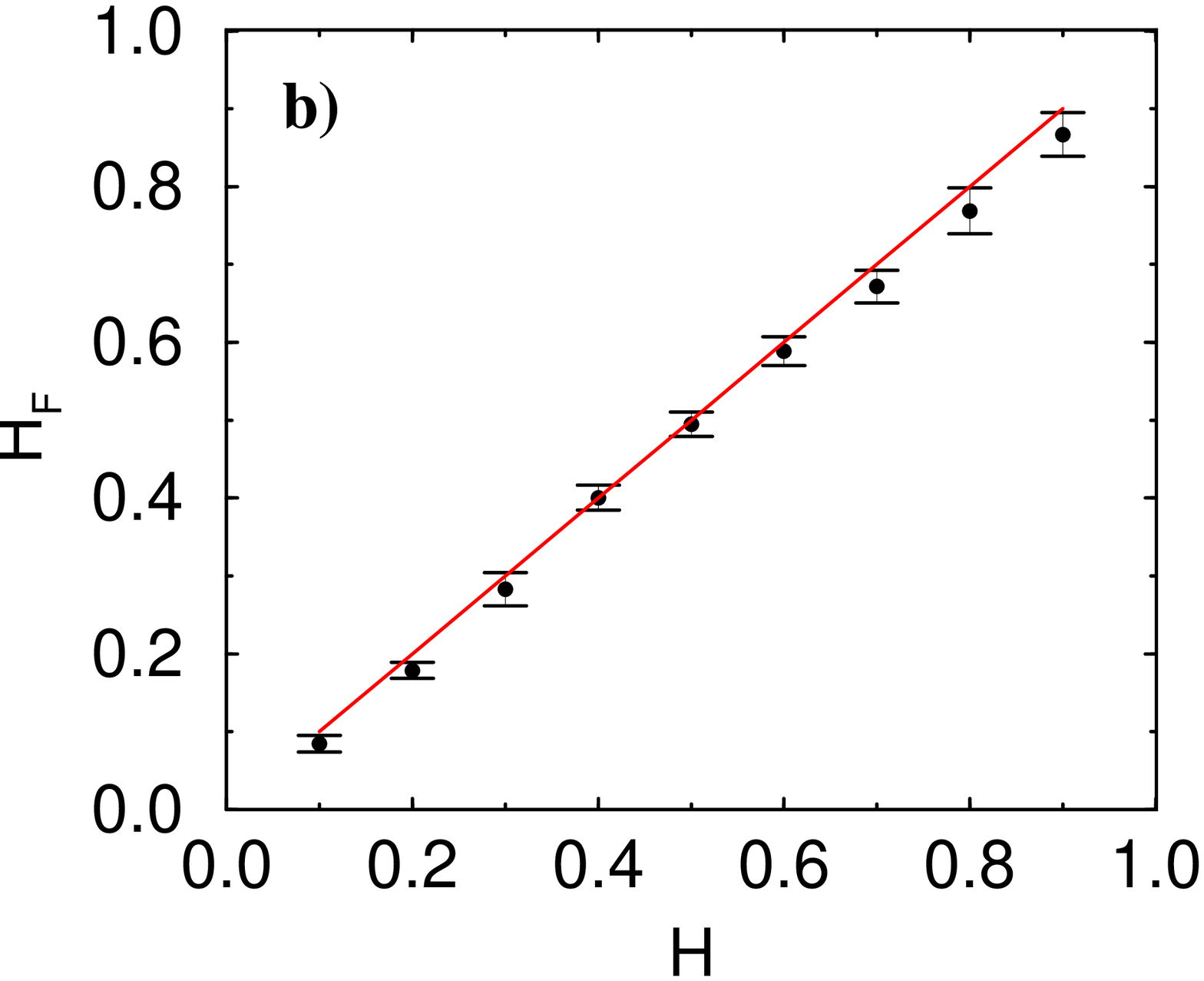,width=8.5cm,height=8.5cm} \\
            \epsfig{file=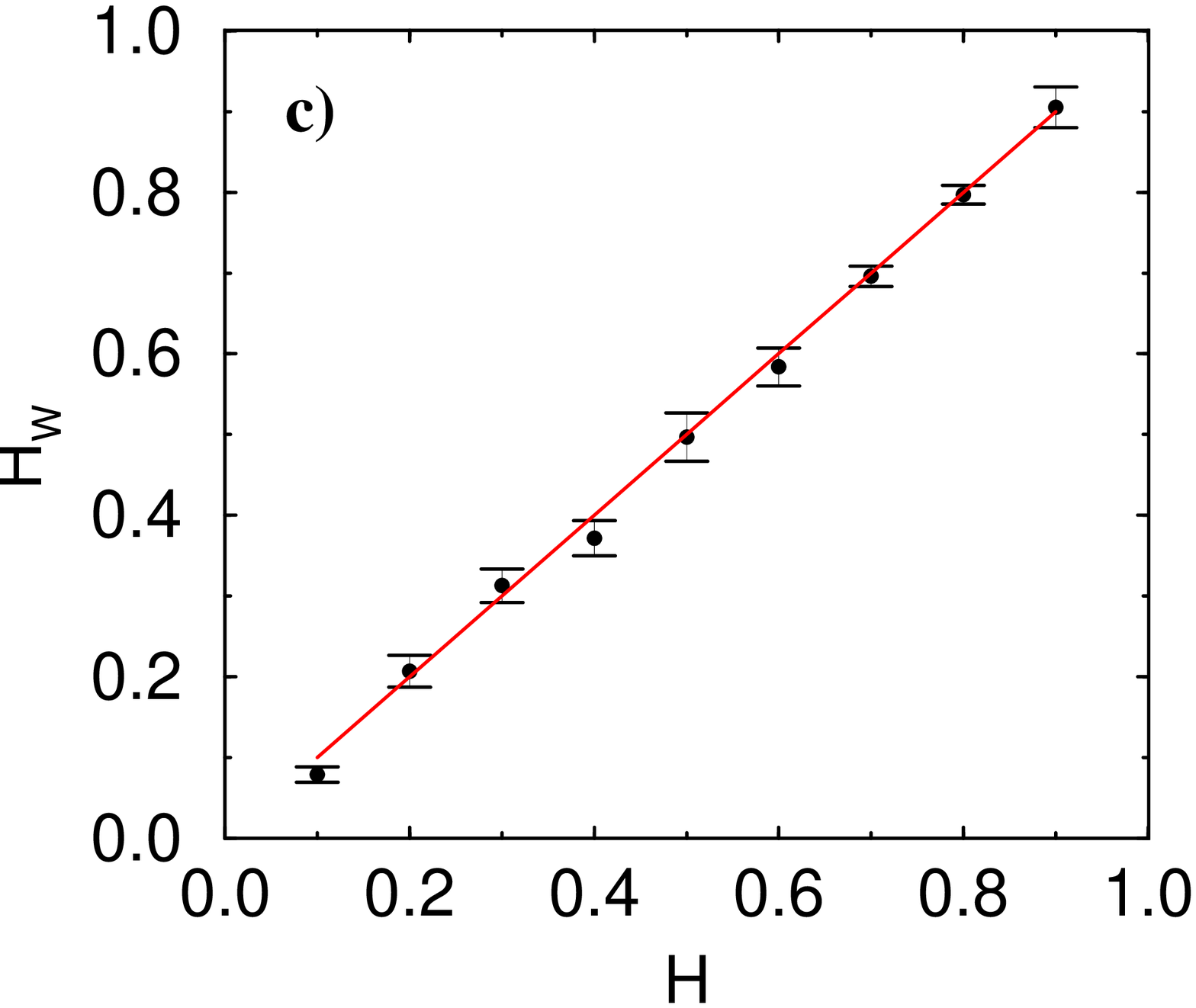,width=8.5cm,height=8.5cm}  &
            \epsfig{file=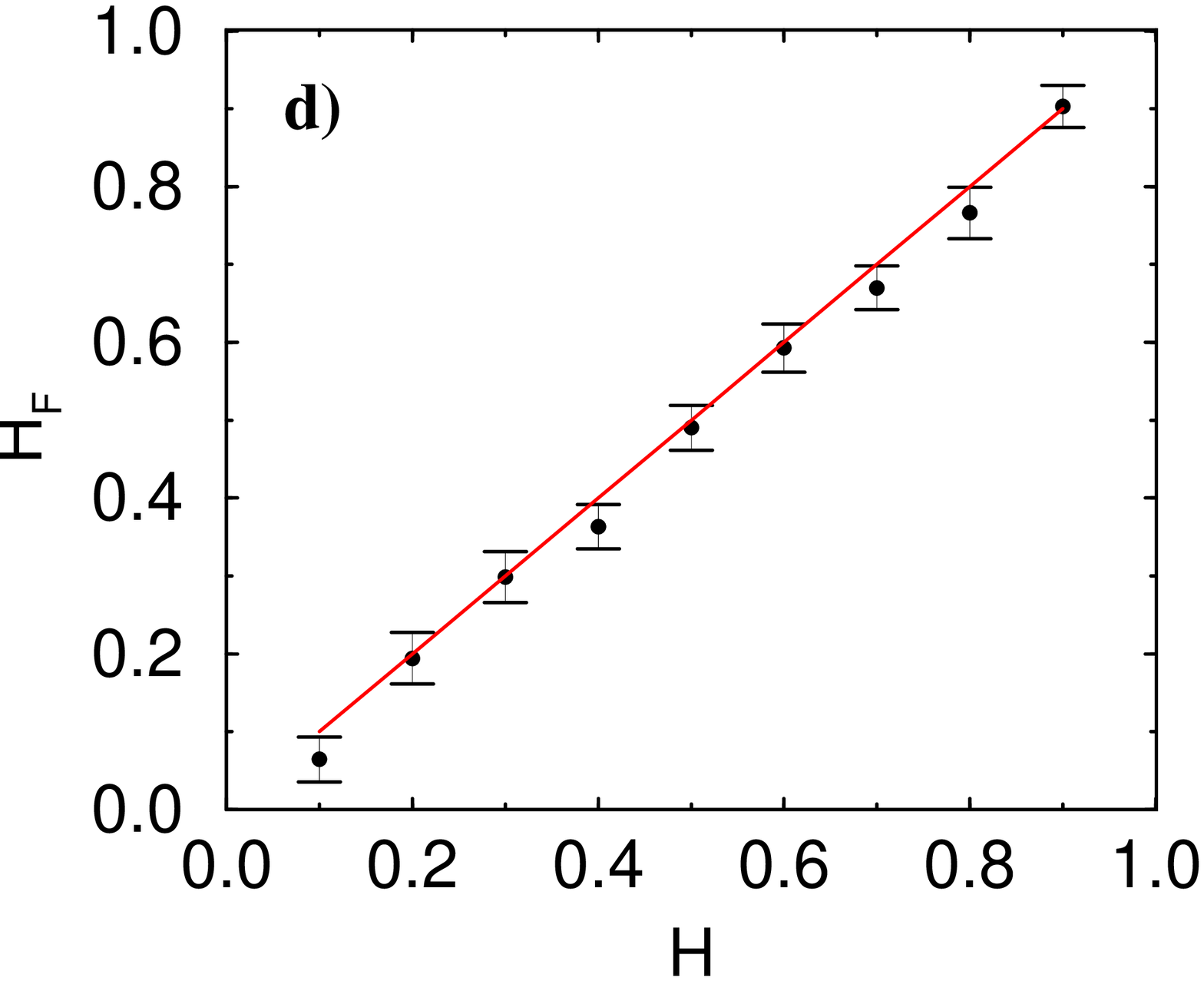,width=8.5cm,height=8.5cm} 
        \end{tabular}
    \end{center}
    \mycaption{\myauthor}{\mytitle}
\end{figure}

\begin{figure}
    \begin{center}
        \begin{tabular}{@{}c@{\hspace{1.0cm}}c@{}}
            \epsfig{file=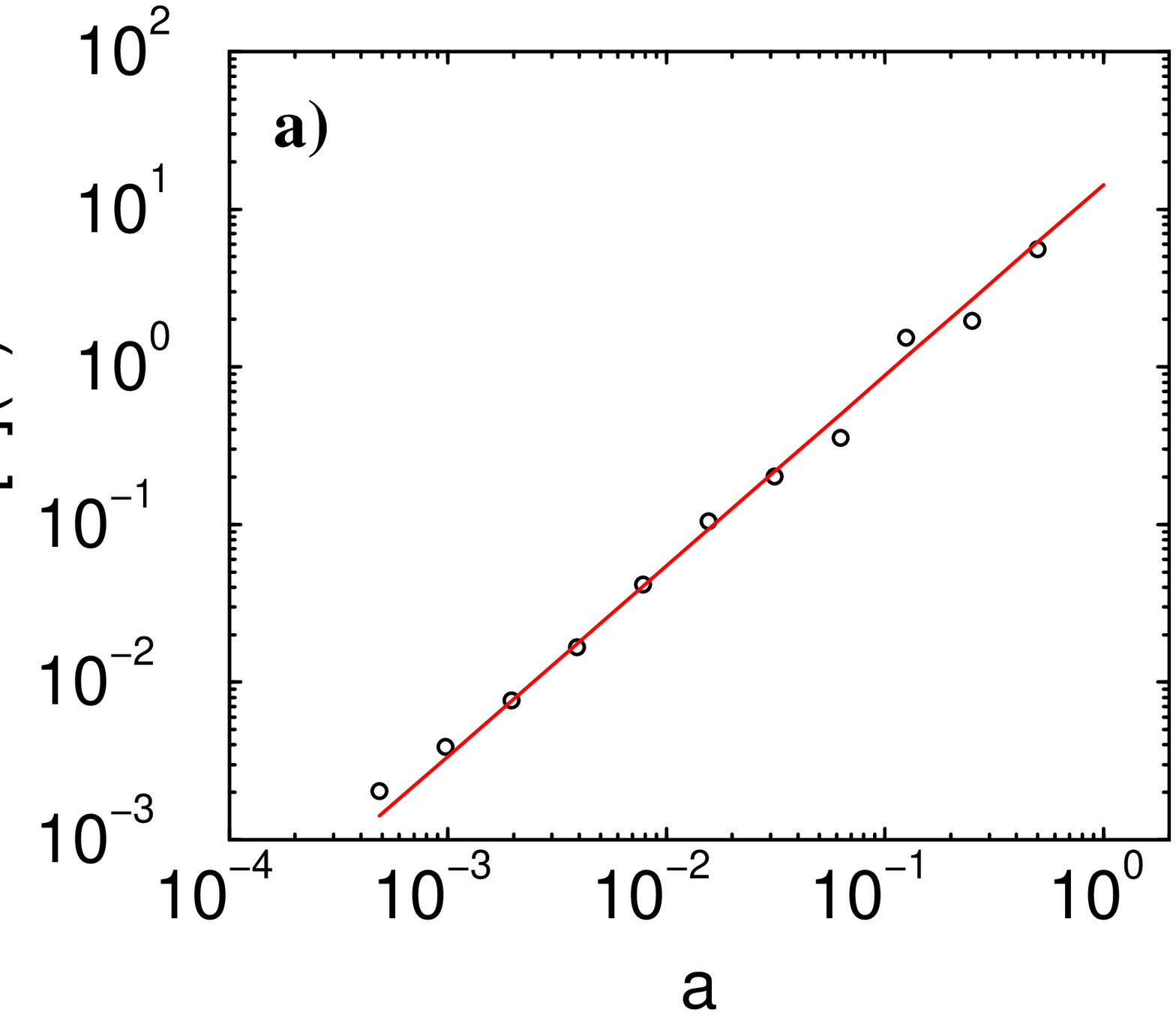,width=8.5cm,height=8.5cm} &
            \epsfig{file=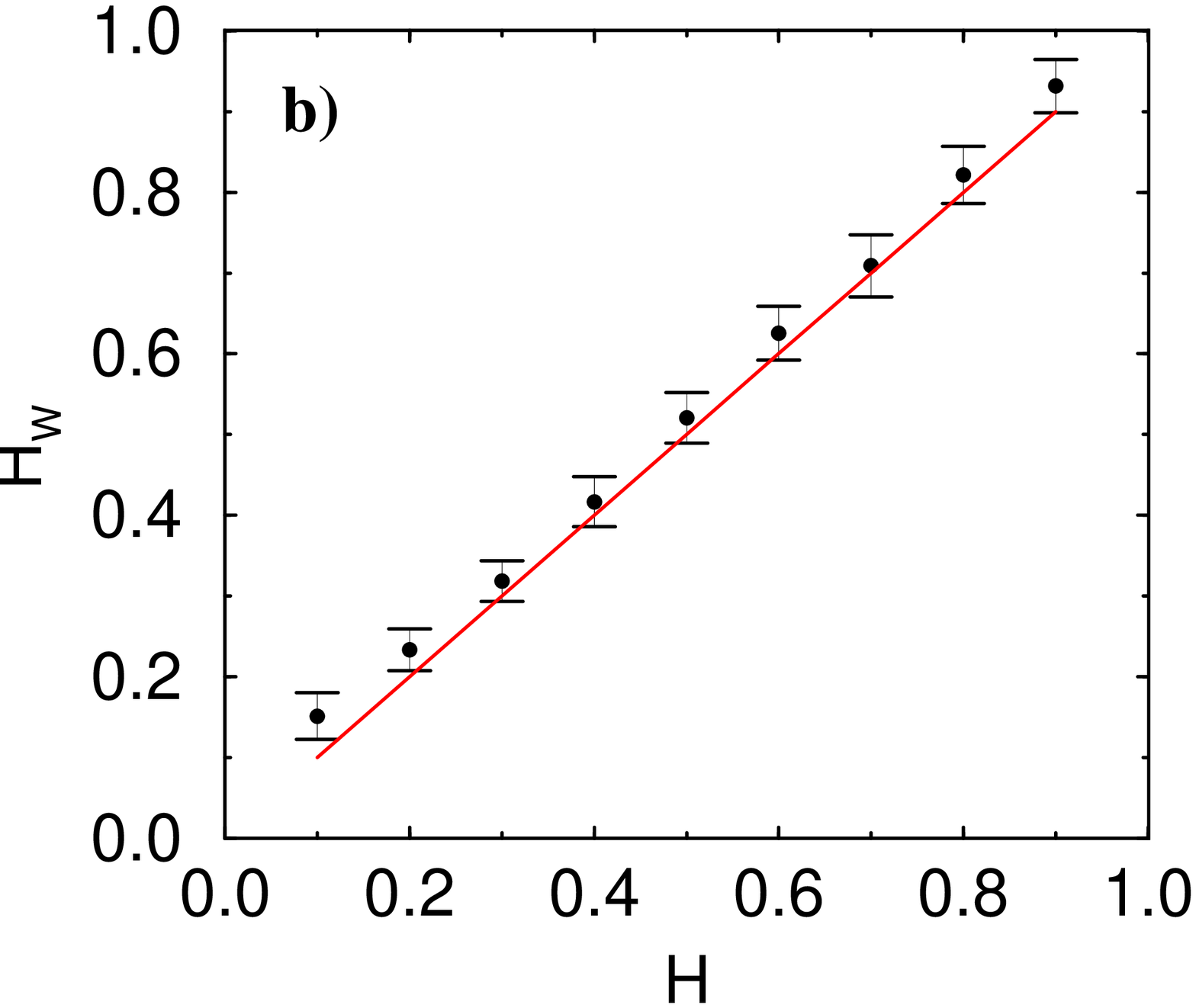,width=8.5cm,height=8.5cm}\\
            \epsfig{file=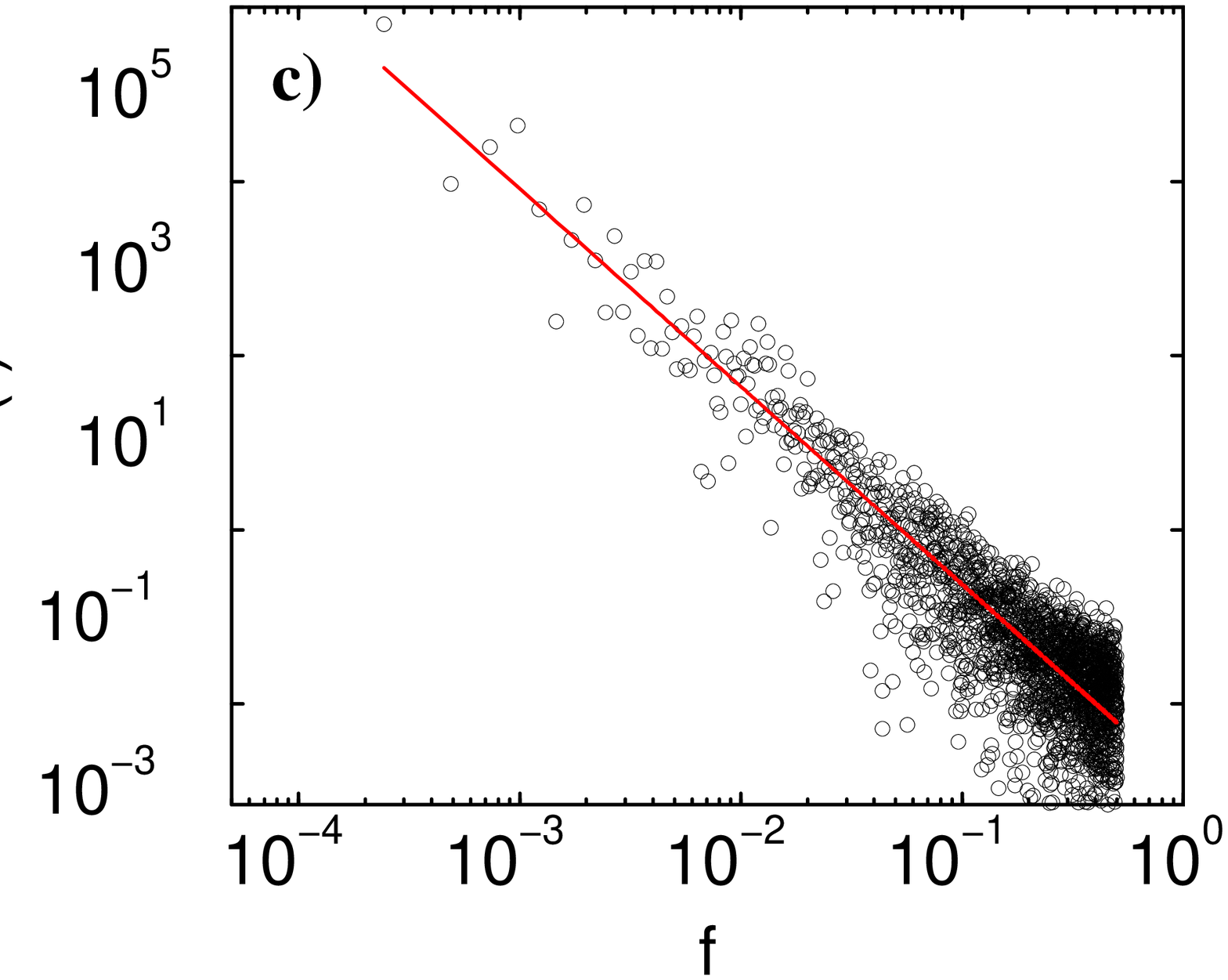,width=8.5cm,height=8.5cm} &
            \epsfig{file=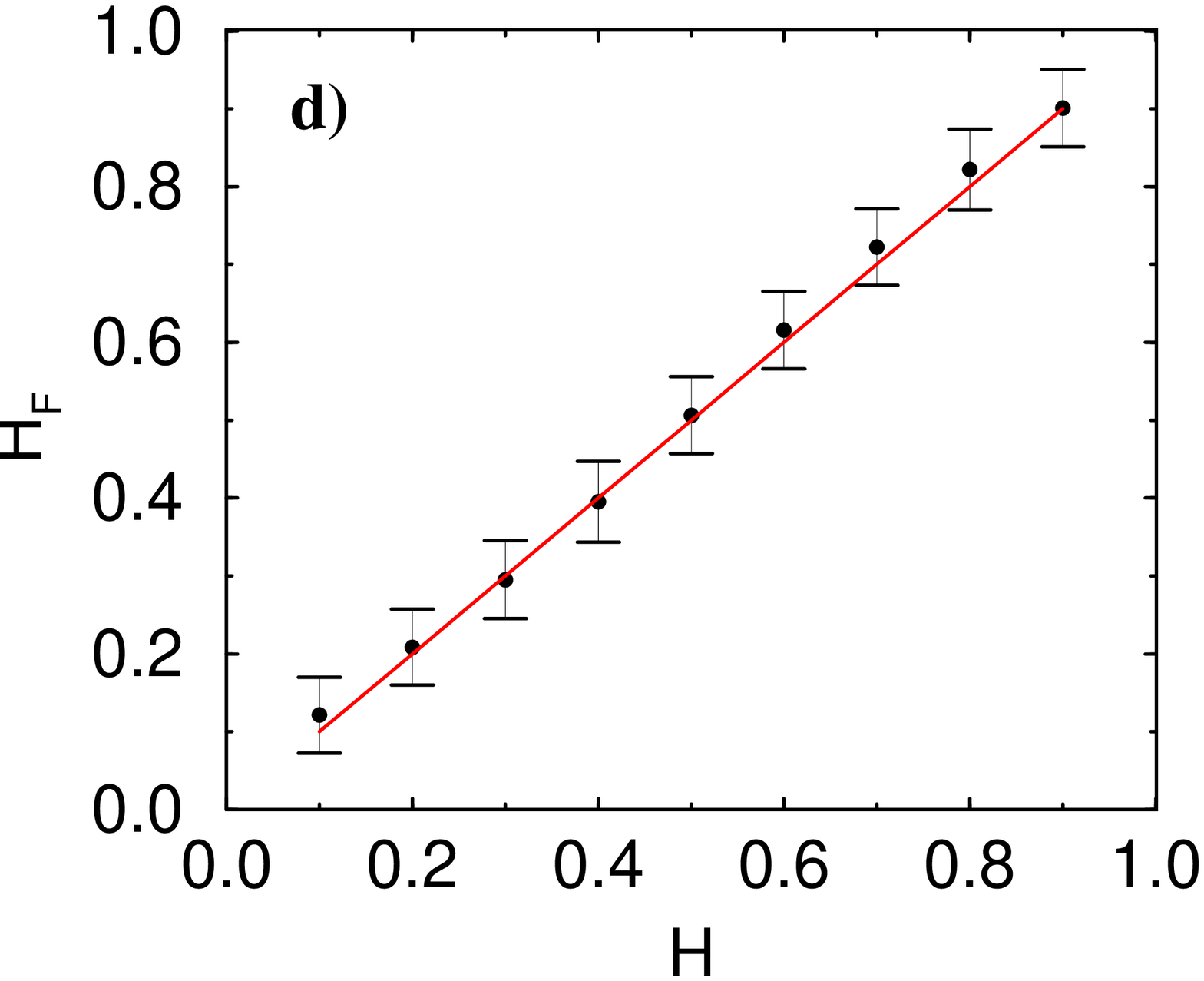,width=8.5cm,height=8.5cm} 
        \end{tabular}
    \end{center}
    \mycaption{\myauthor}{\mytitle}
\end{figure}

\begin{figure}
    \begin{center}
        \epsfig{file=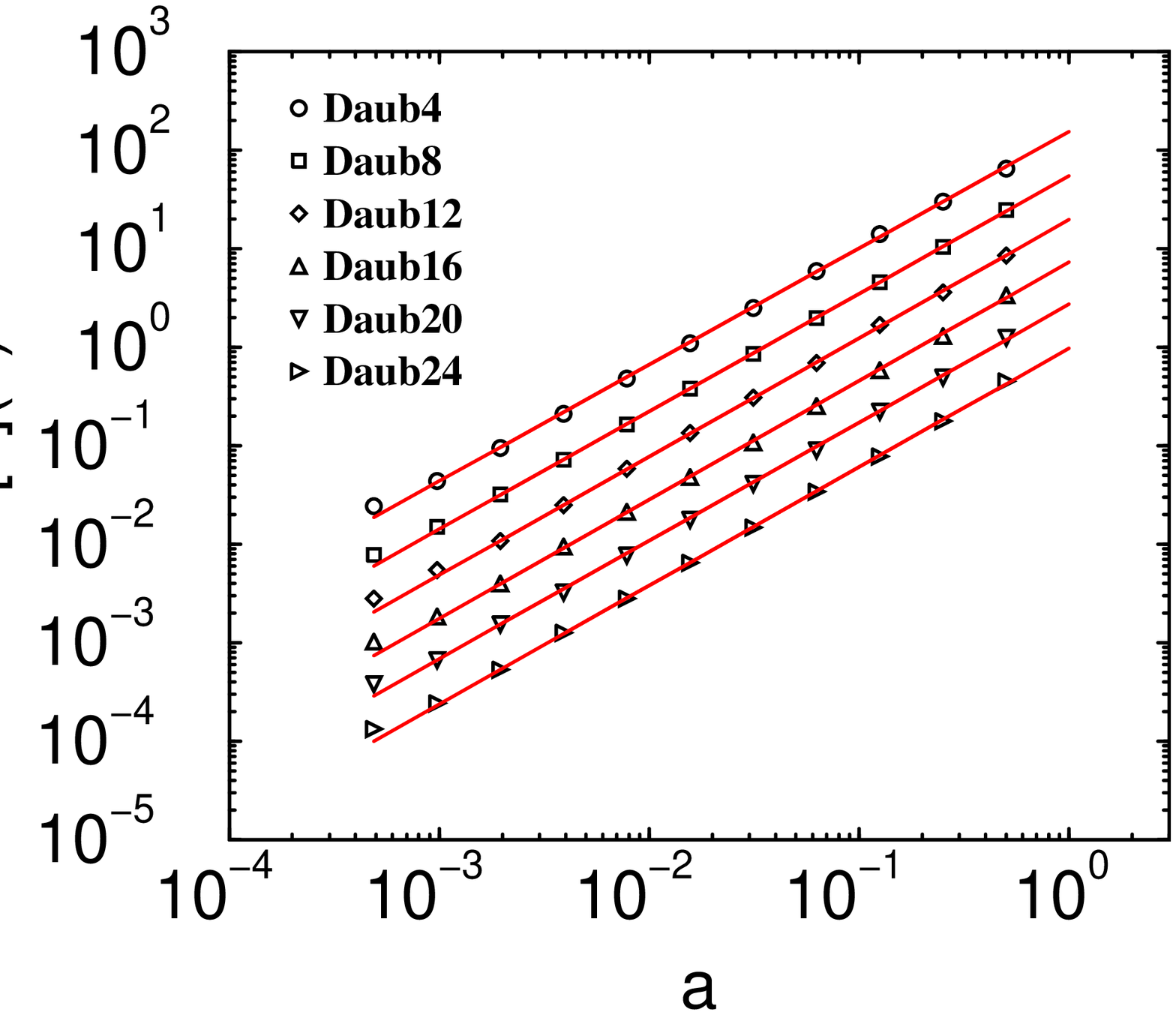,width=8.5cm,height=8.5cm}
    \end{center}
    \mycaption{\myauthor}{\mytitle}
\newpage 

    \begin{center}
        \begin{tabular}{@{}c@{\hspace{1.0cm}}c@{}}
            \epsfig{file=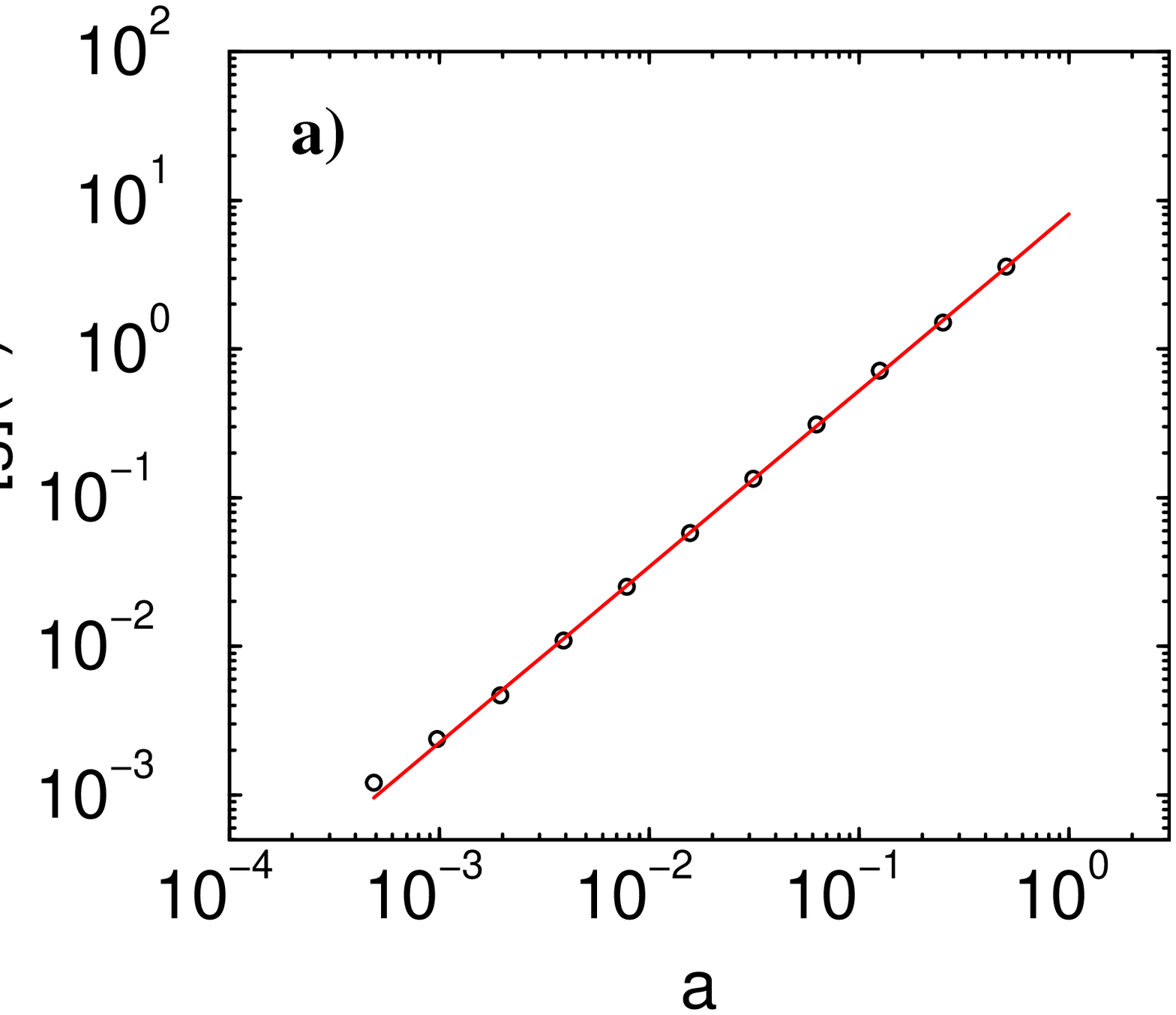,width=8.5cm,height=8.5cm} &
            \epsfig{file=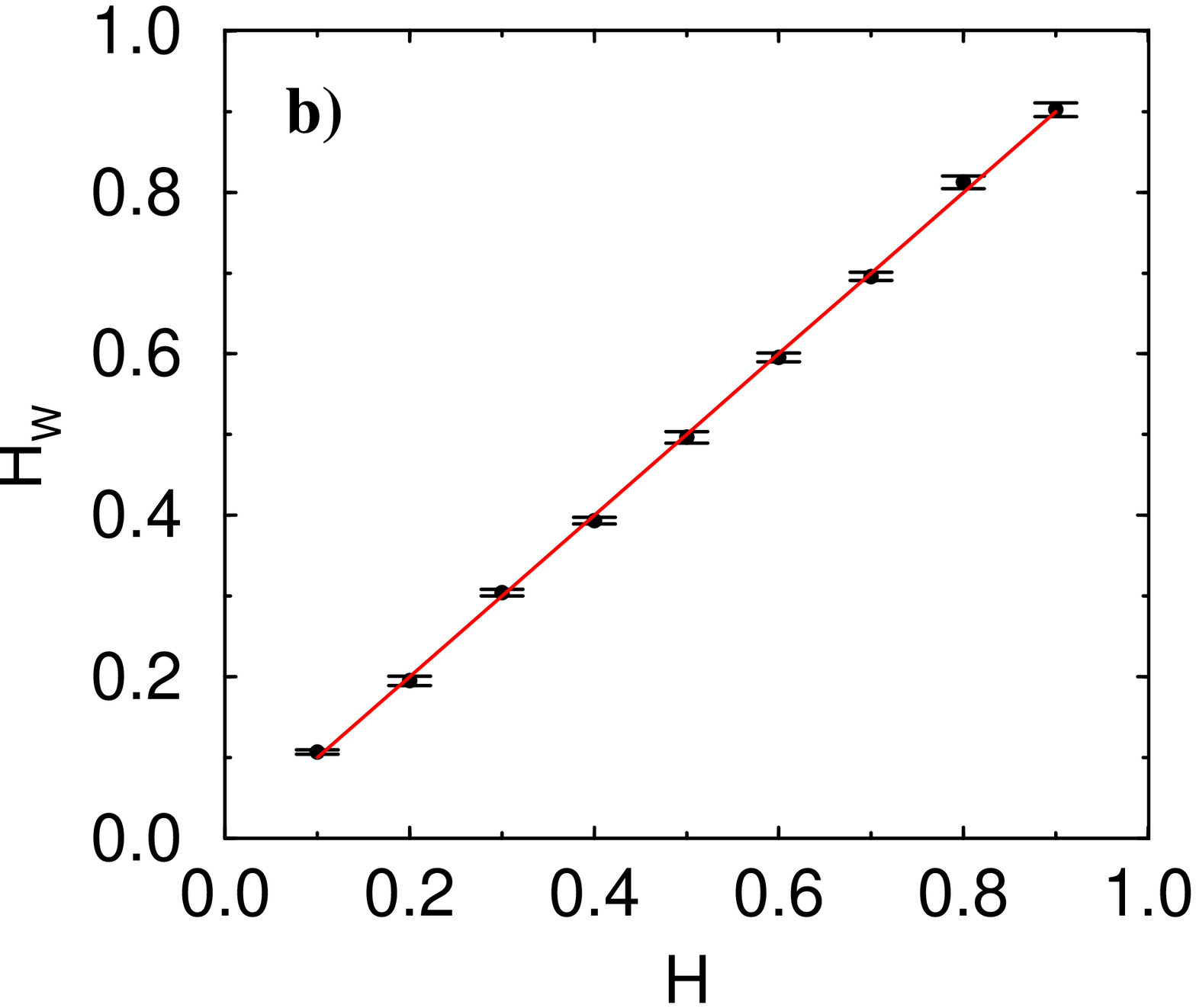,width=8.5cm,height=8.5cm} 
        \end{tabular}
    \end{center}
    \mycaption{\myauthor}{\mytitle}
\end{figure}

\begin{figure}
    \begin{center}
        \begin{tabular}{@{}c@{\hspace{1.0cm}}c@{}}
            \epsfig{file=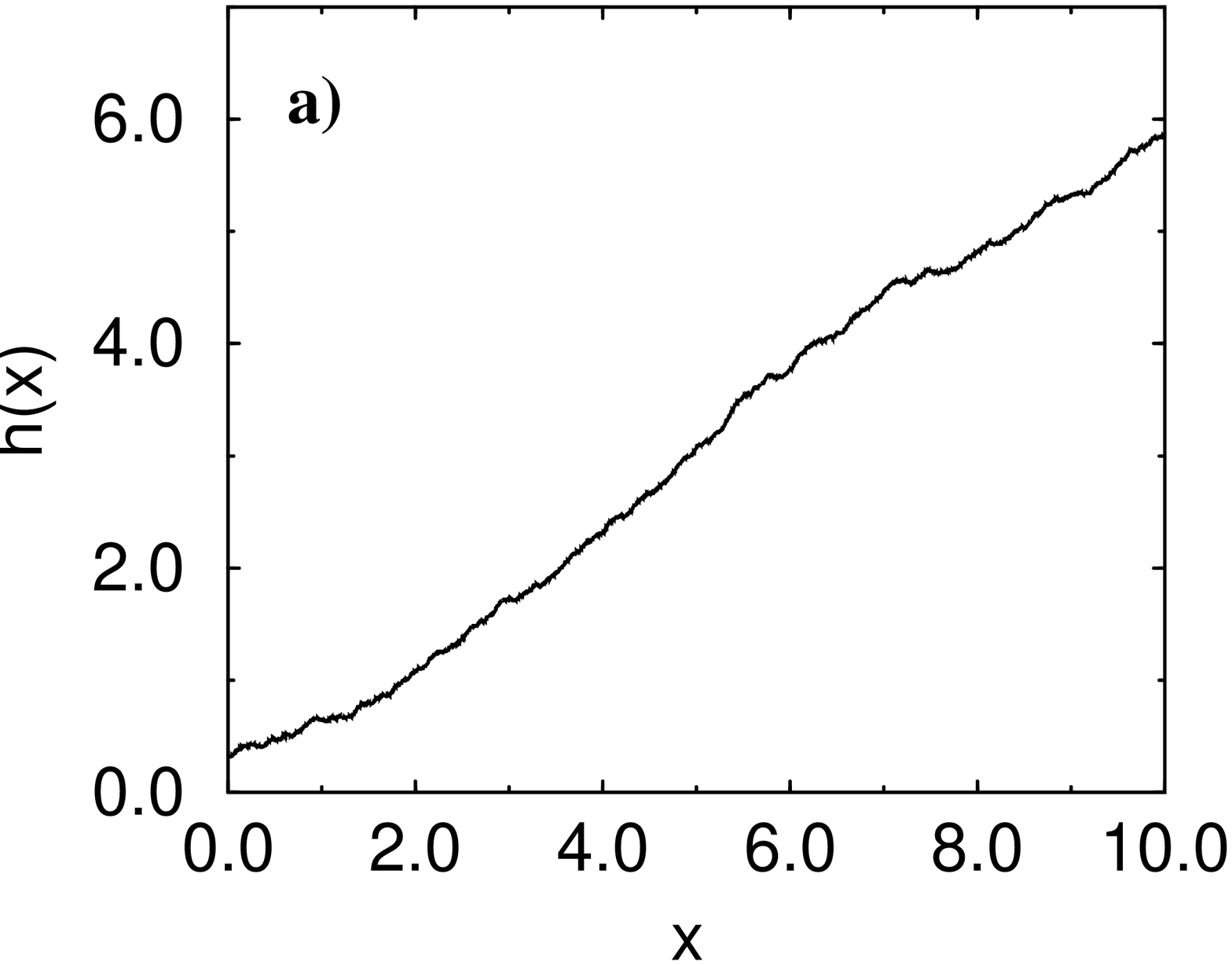,width=8.5cm,height=8.5cm} &
            \epsfig{file=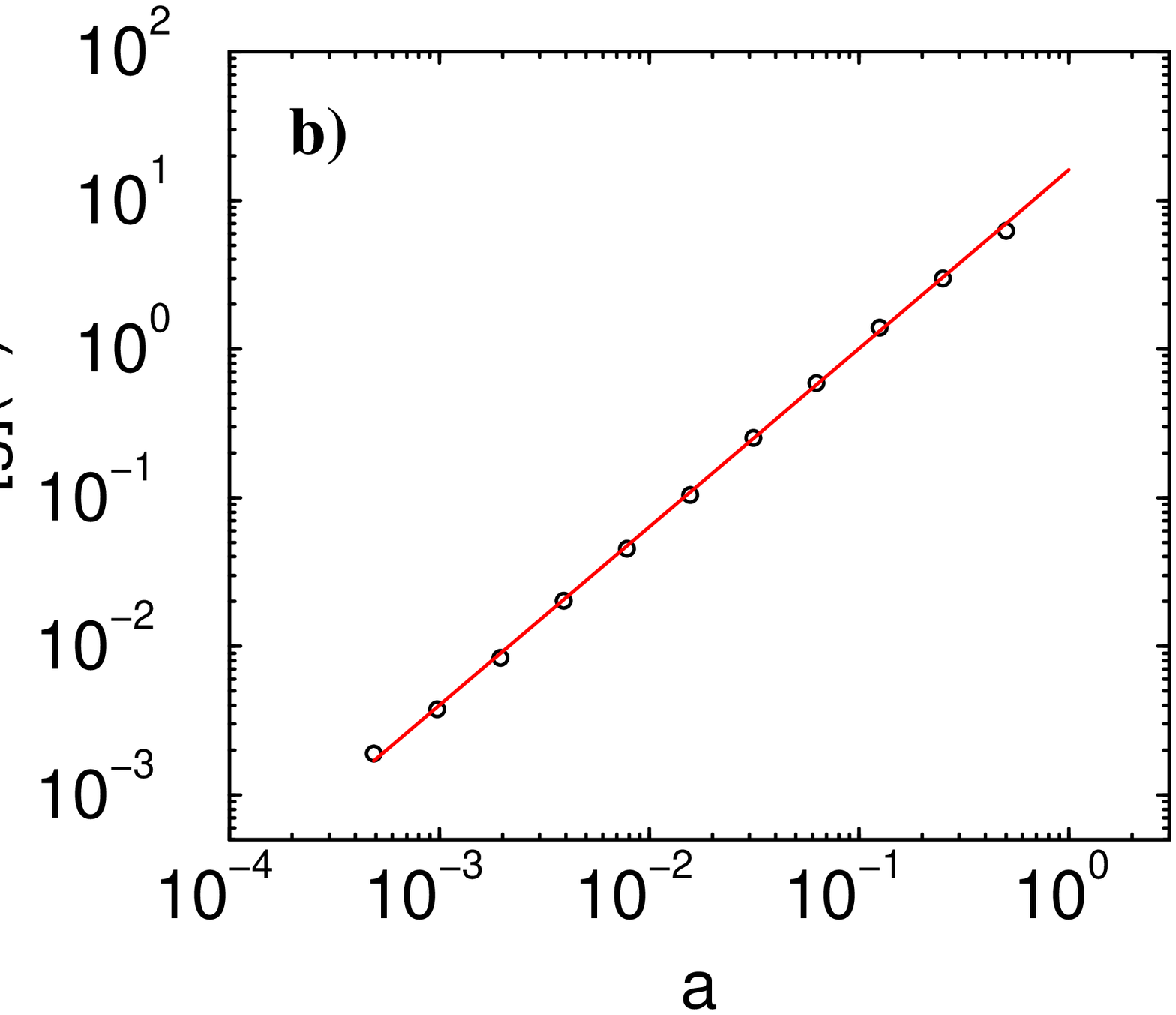,width=8.5cm,height=8.5cm} \\
            \epsfig{file=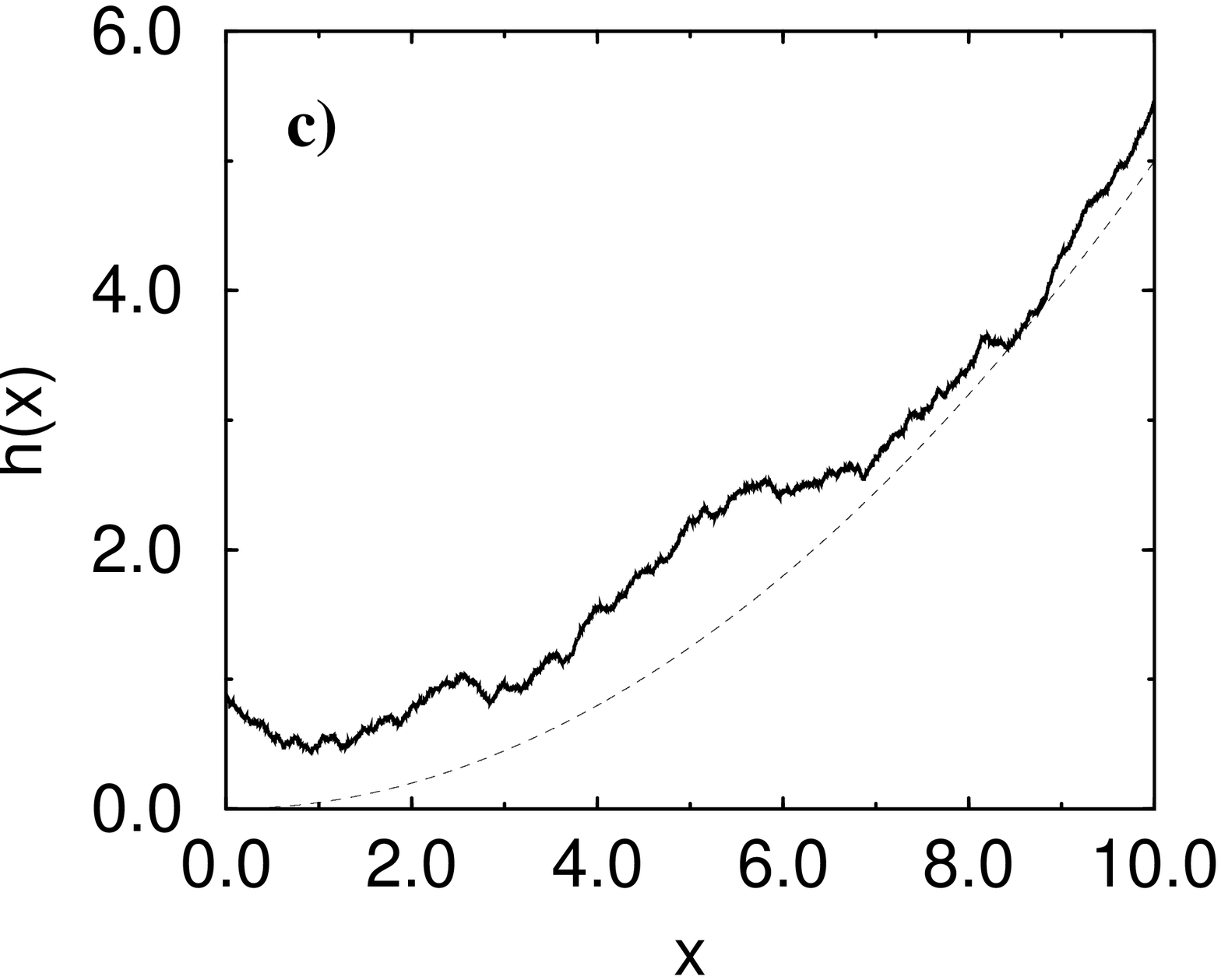,width=8.5cm,height=8.5cm} &
            \epsfig{file=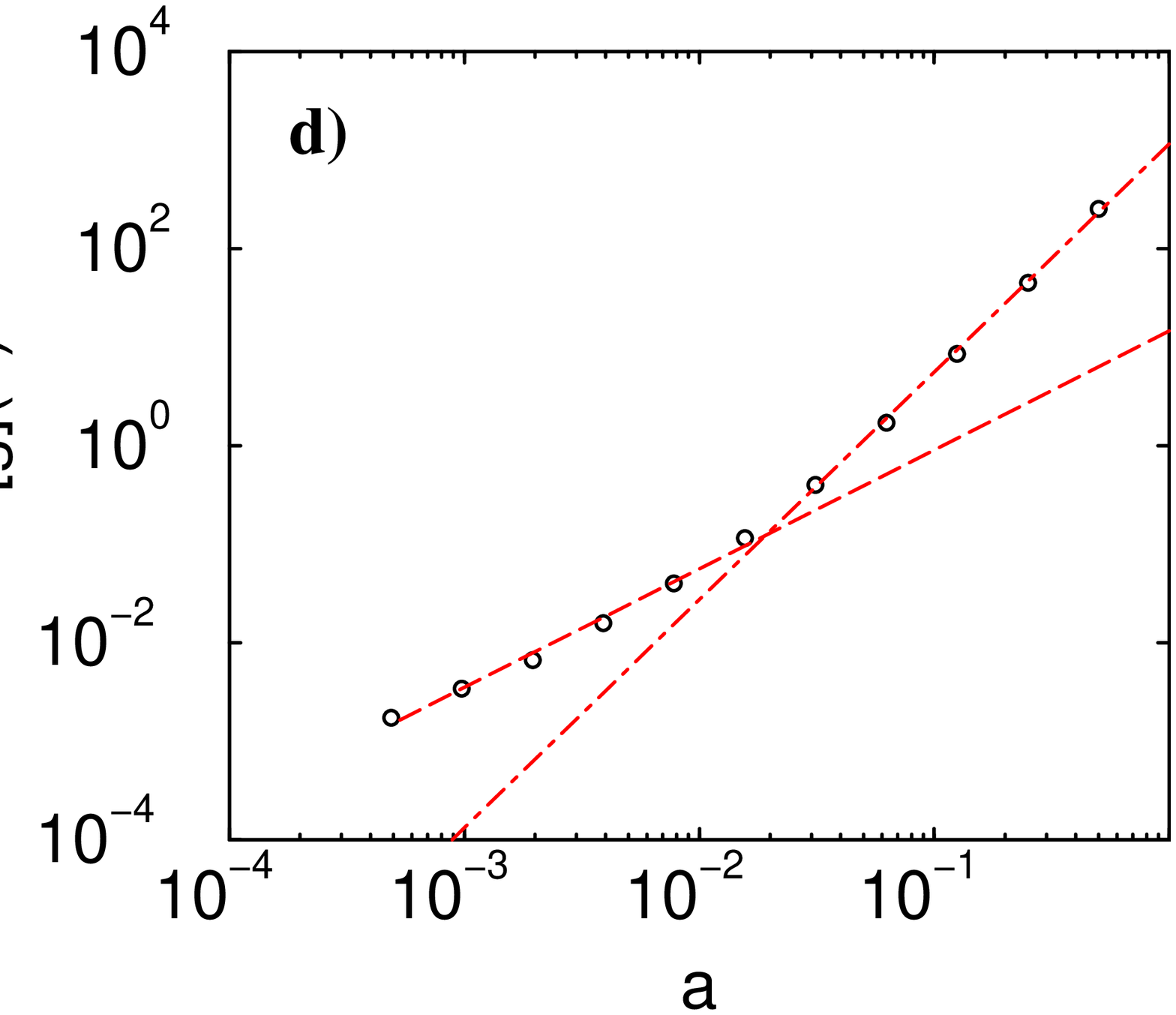,width=8.5cm,height=8.5cm} 
        \end{tabular}
    \end{center}
    \mycaption{\myauthor}{\mytitle}
\end{figure}

\newpage

\begin{figure}
    \begin{center}
        \begin{tabular}{@{}c@{\hspace{1.0cm}}c@{}}
            \epsfig{file=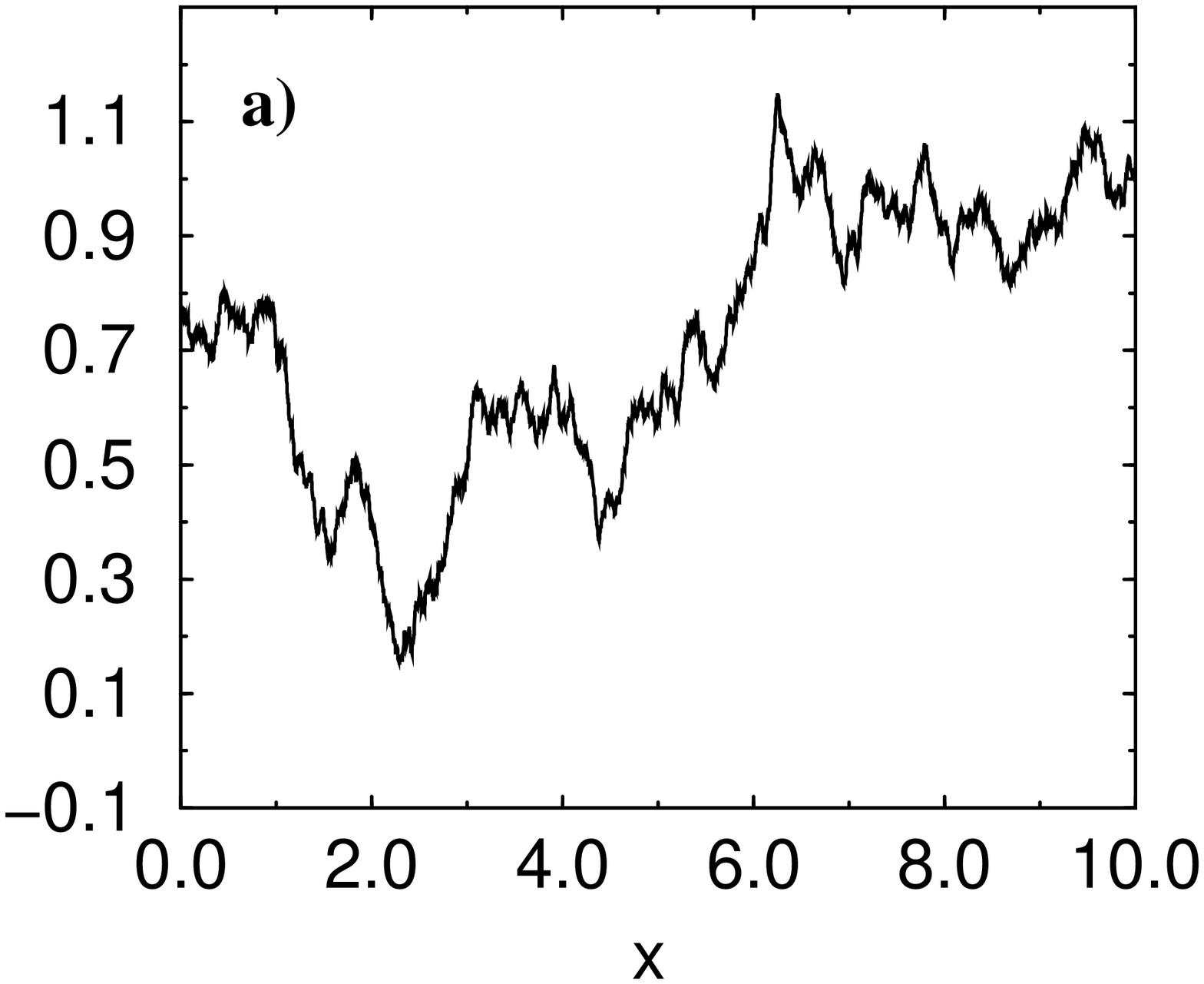,width=8.5cm,height=8.5cm} &
            \epsfig{file=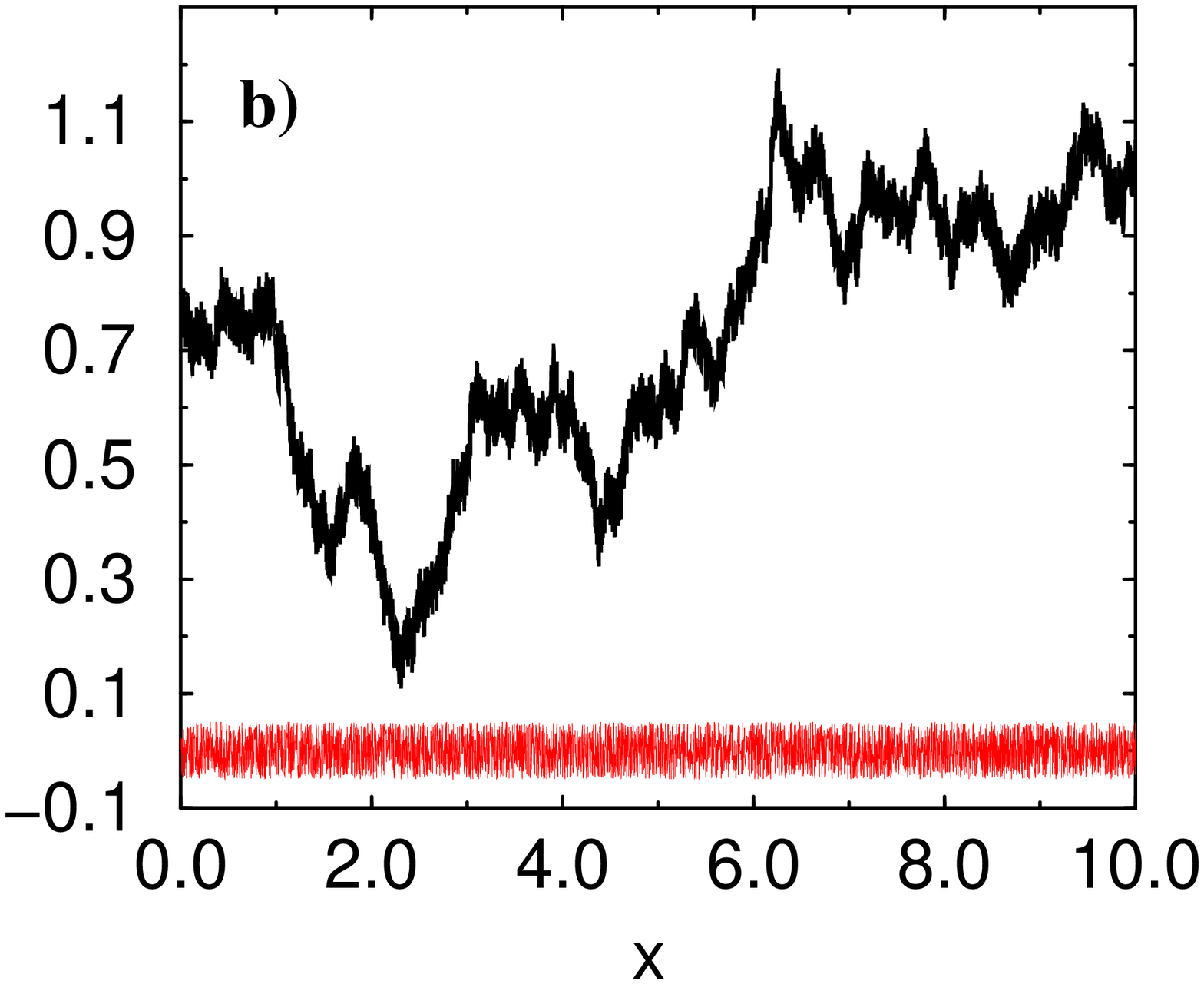,width=8.5cm,height=8.5cm} \\
            \epsfig{file=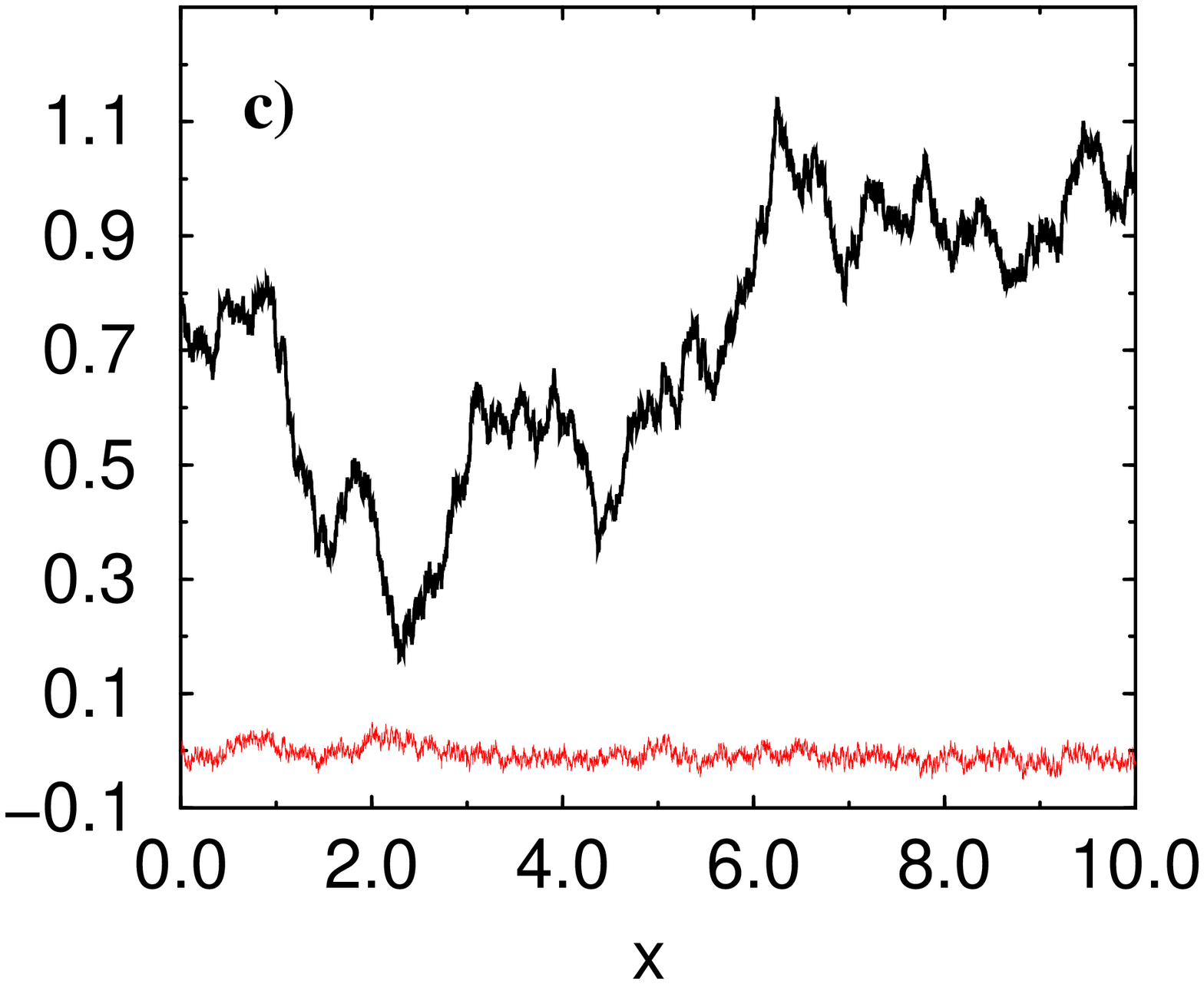,width=8.5cm,height=8.5cm} &
            \epsfig{file=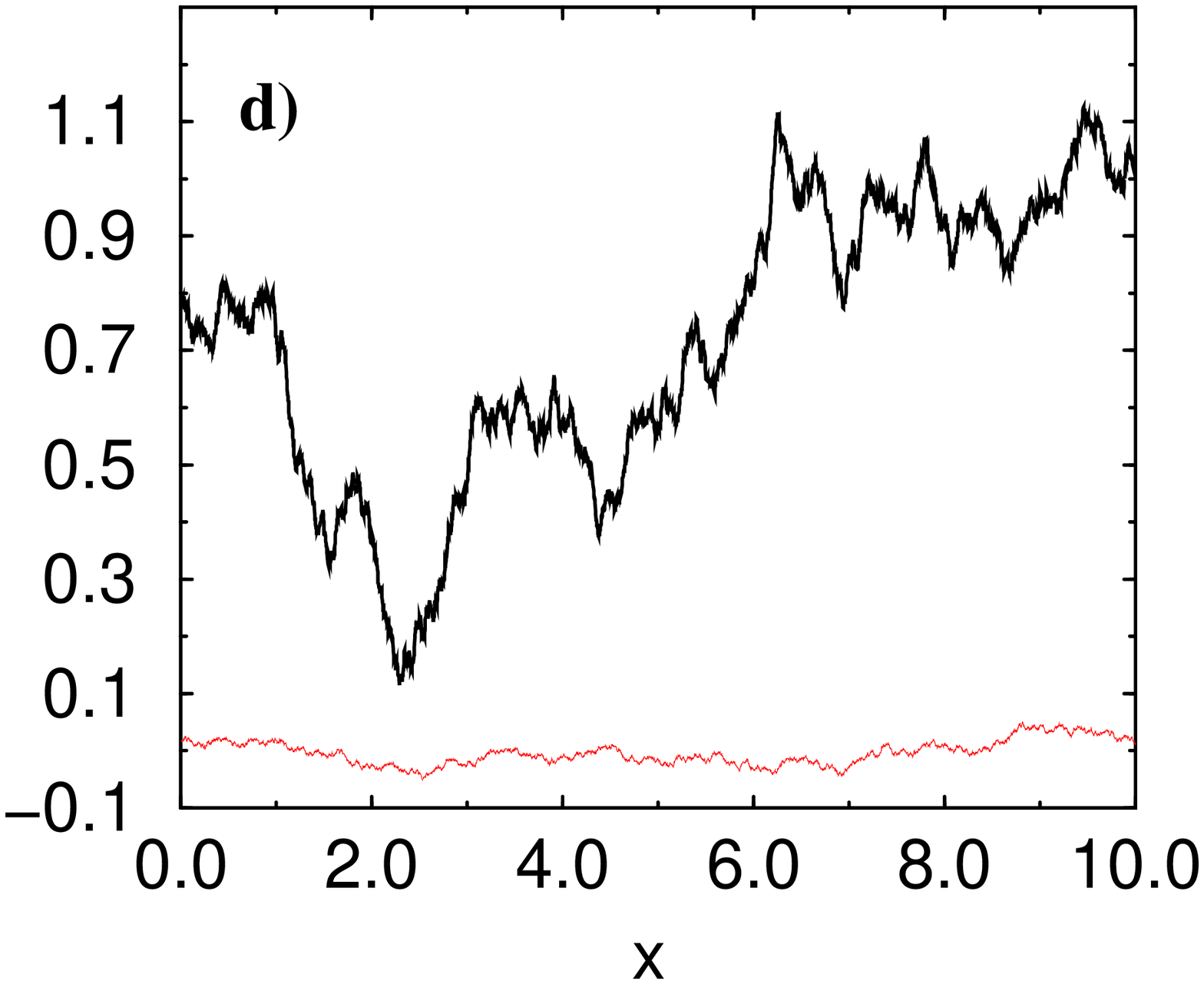,width=8.5cm,height=8.5cm} 
        \end{tabular}
    \end{center}
    \mycaption{\myauthor}{\mytitle}
\end{figure}

\begin{figure}
    \begin{center}
        \begin{tabular}{@{}c@{\hspace{1.0cm}}c@{}}
            \epsfig{file=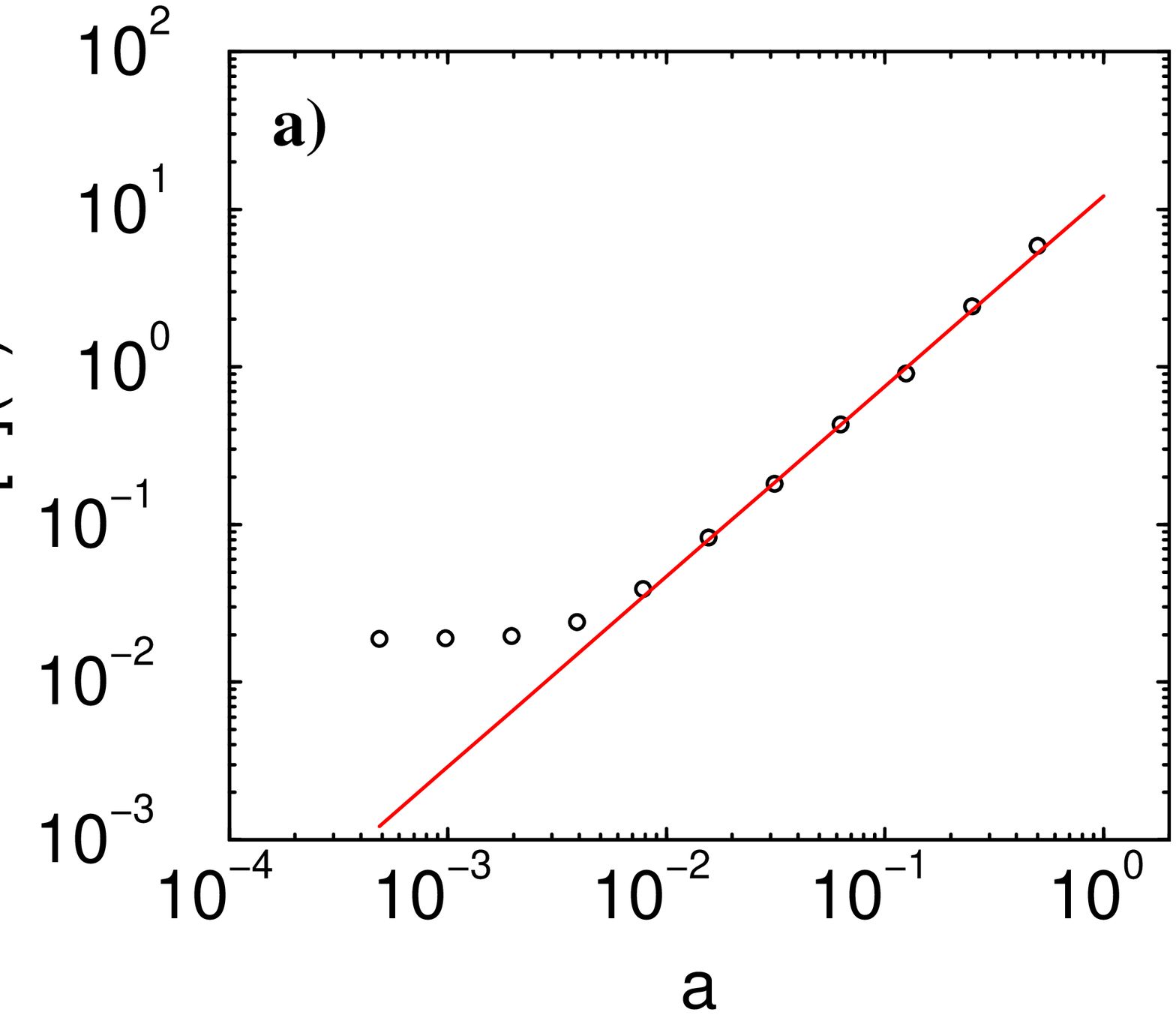,width=8.5cm,height=8.5cm} &
            \epsfig{file=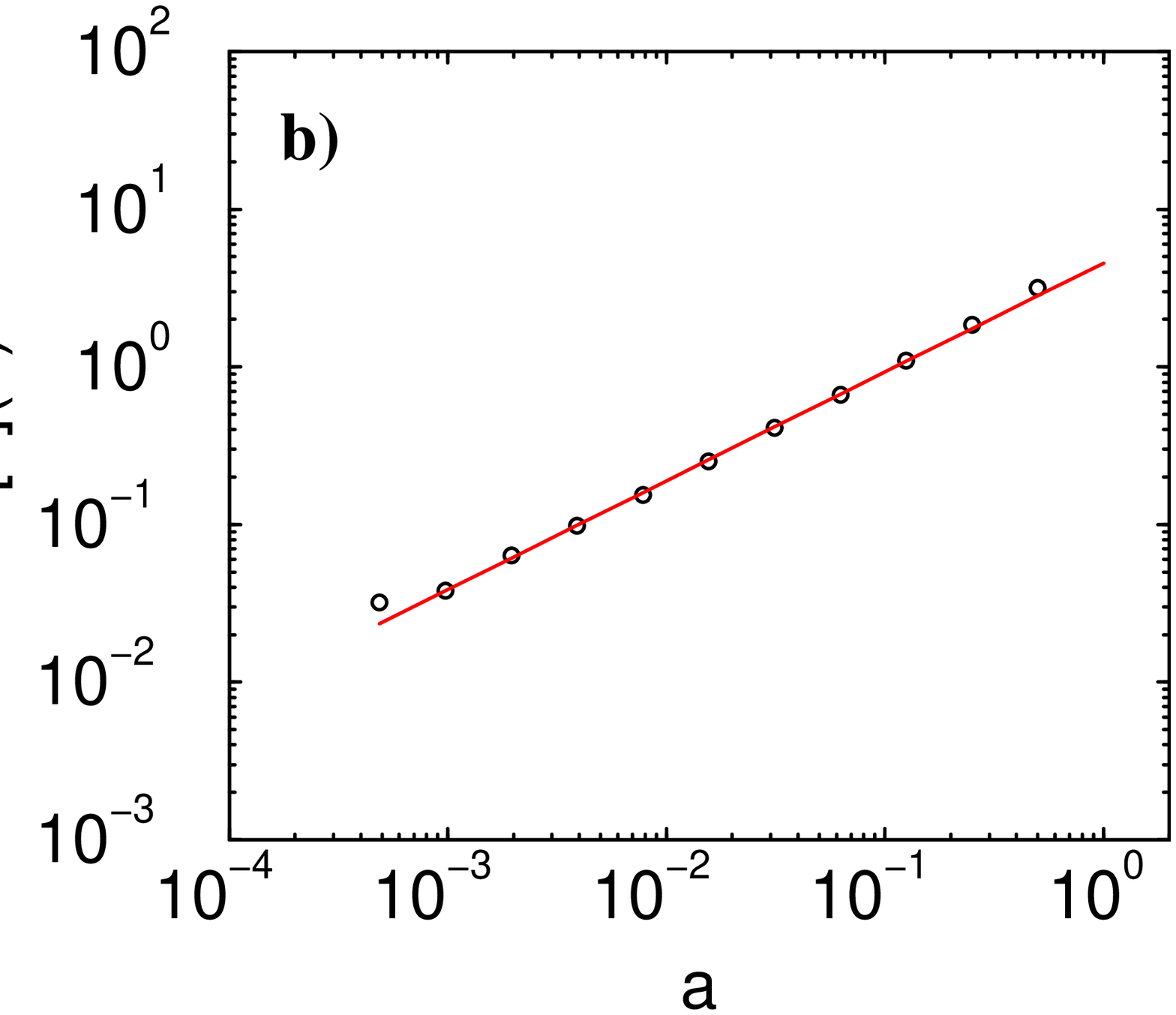,width=8.5cm,height=8.5cm} \\
            \epsfig{file=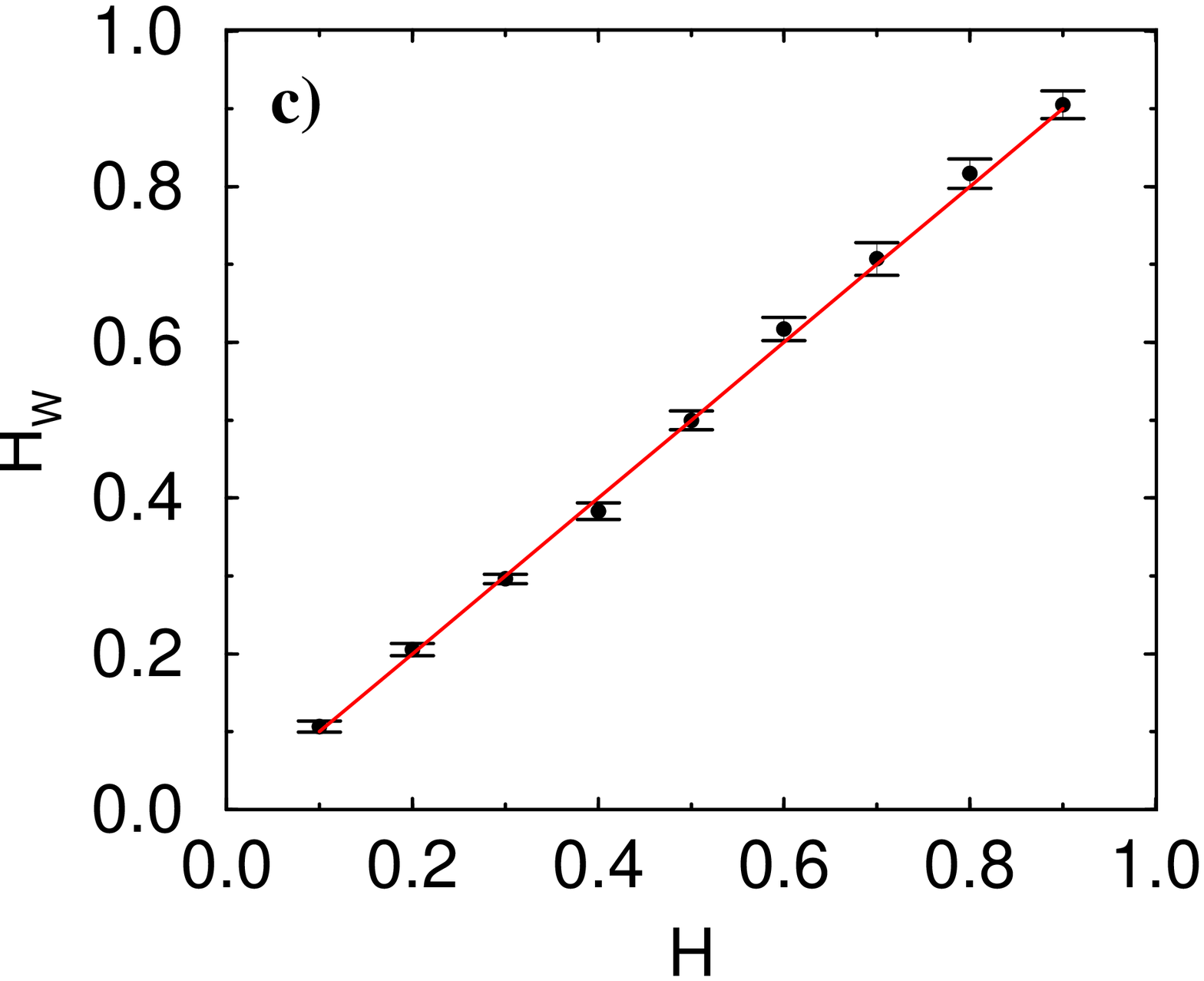,width=8.5cm,height=8.5cm} &
            \epsfig{file=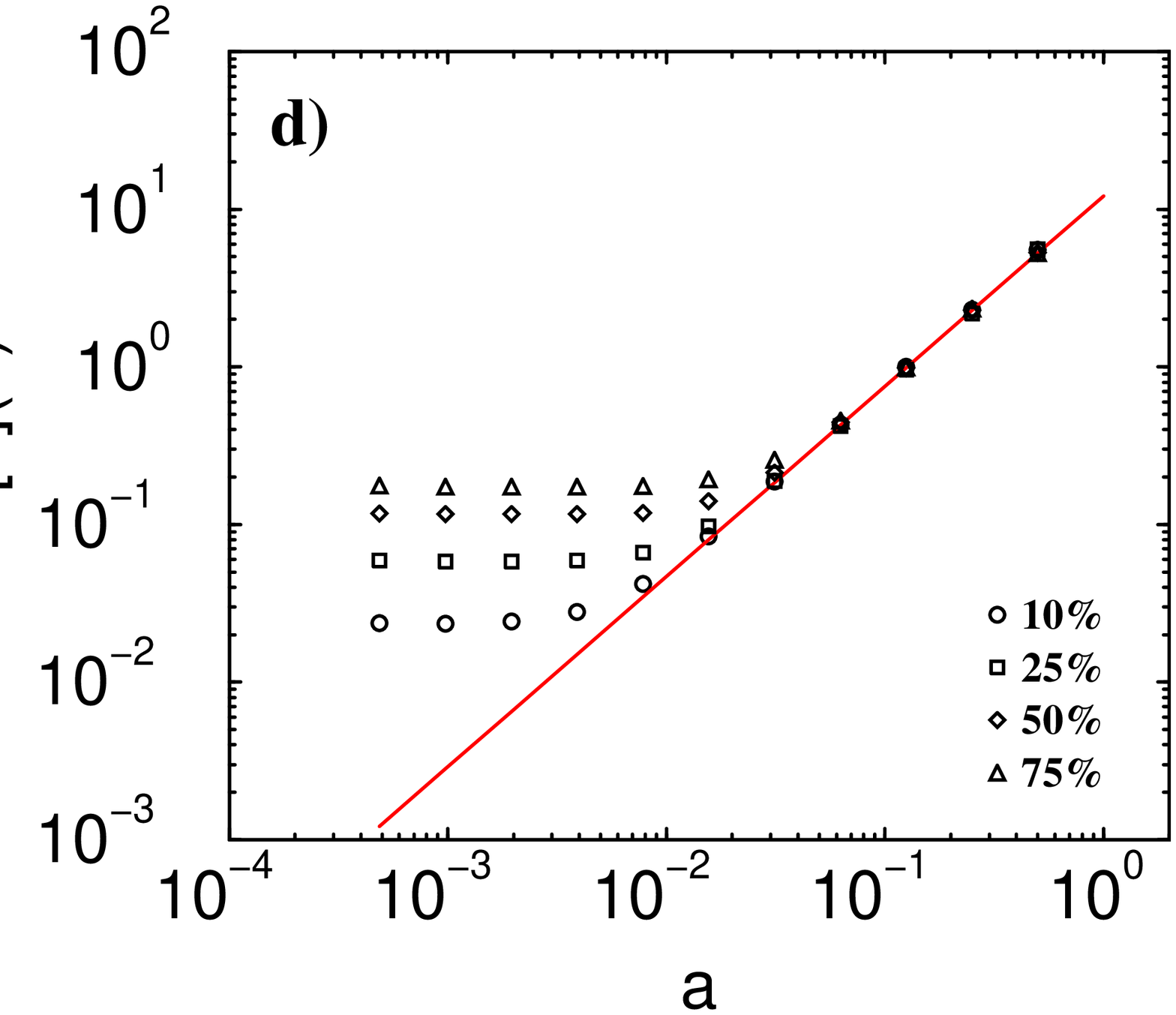,width=8.5cm,height=8.5cm} 
        \end{tabular}
    \end{center}
    \mycaption{\myauthor}{\mytitle}
\end{figure}

\begin{figure}
    \begin{center}
        \begin{tabular}{@{}c@{\hspace{1.0cm}}c@{}}
            \epsfig{file=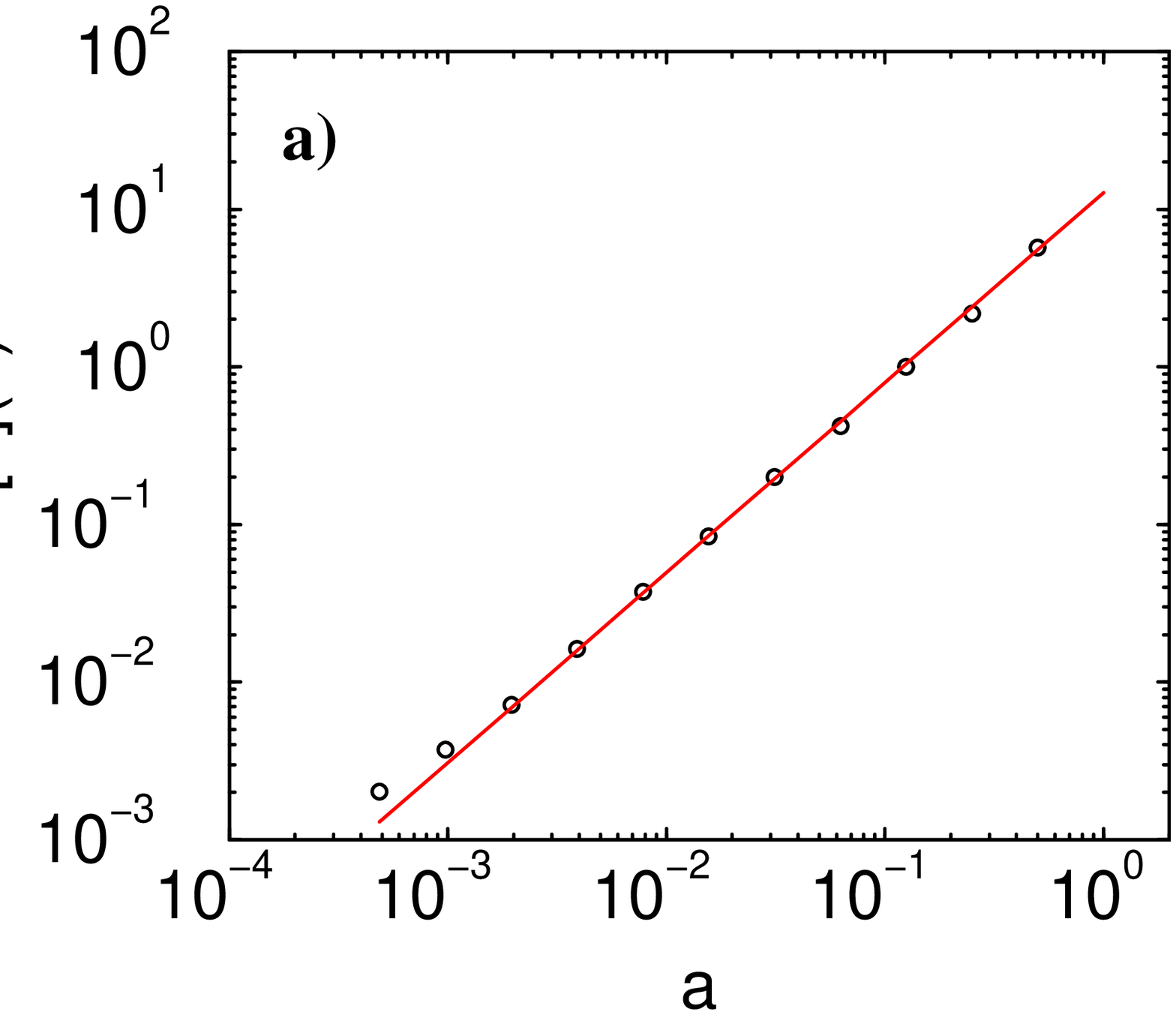,width=8.5cm,height=8.5cm} &
            \epsfig{file=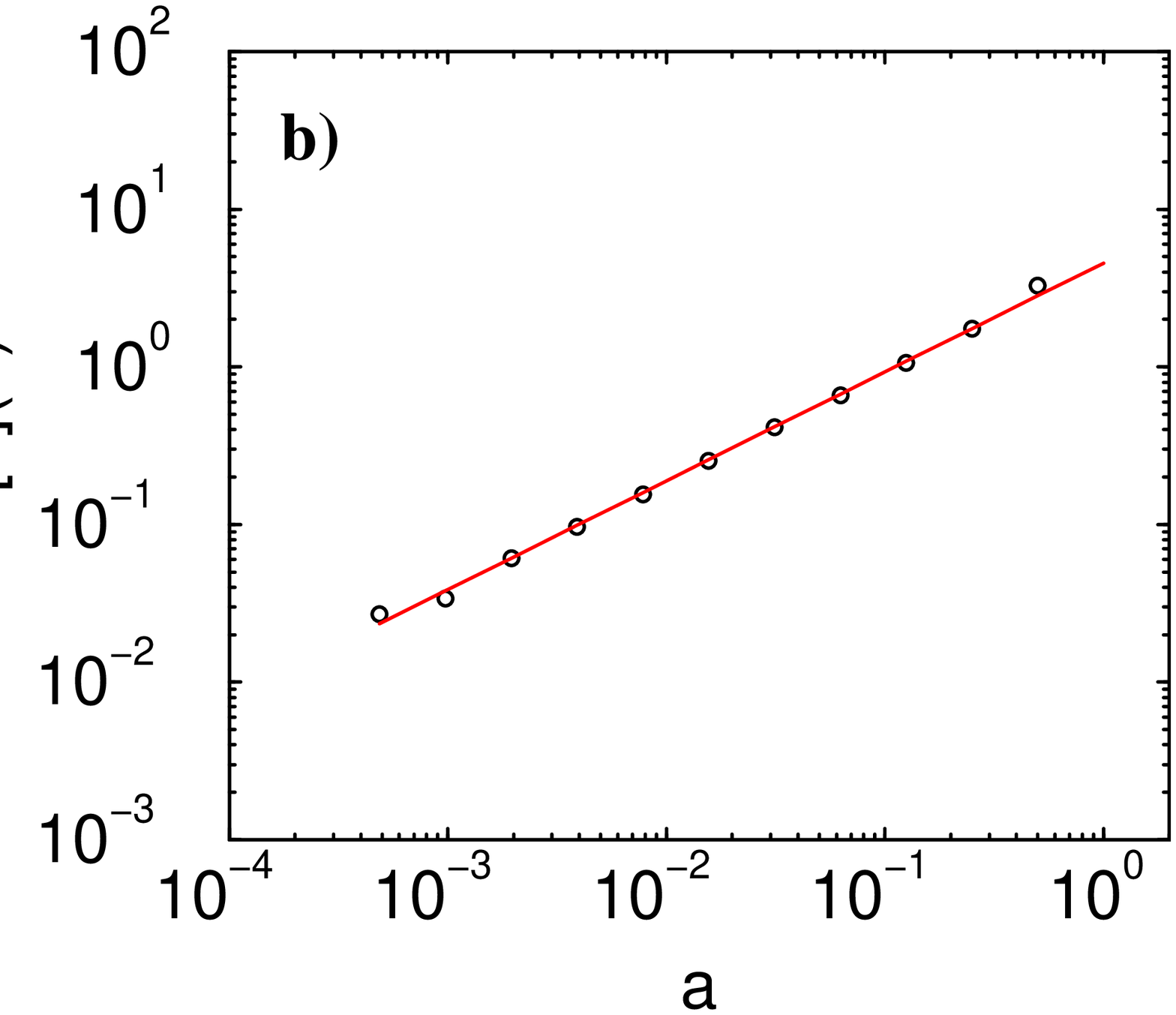,width=8.5cm,height=8.5cm} \\
            \epsfig{file=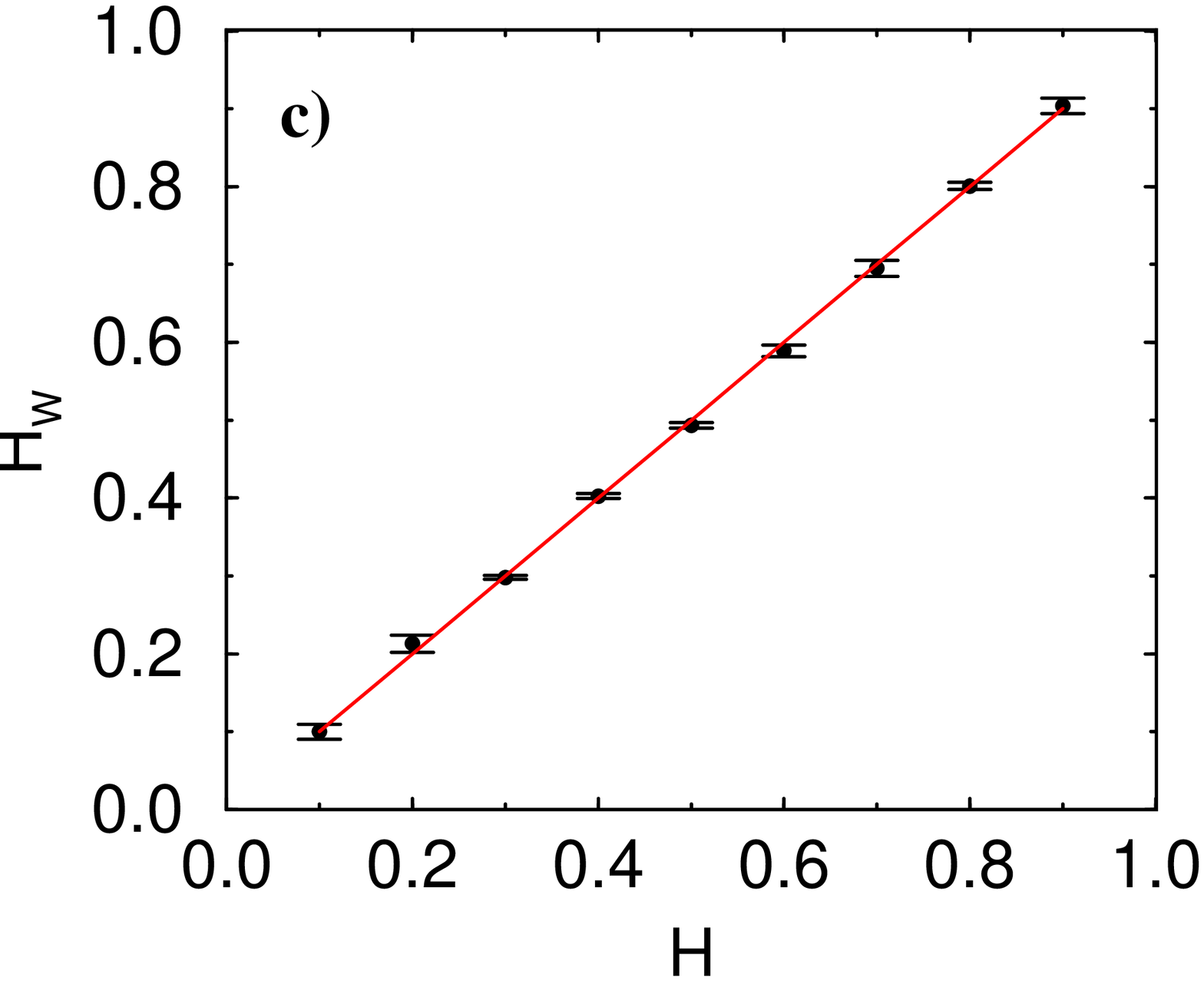,width=8.5cm,height=8.5cm} &
            \epsfig{file=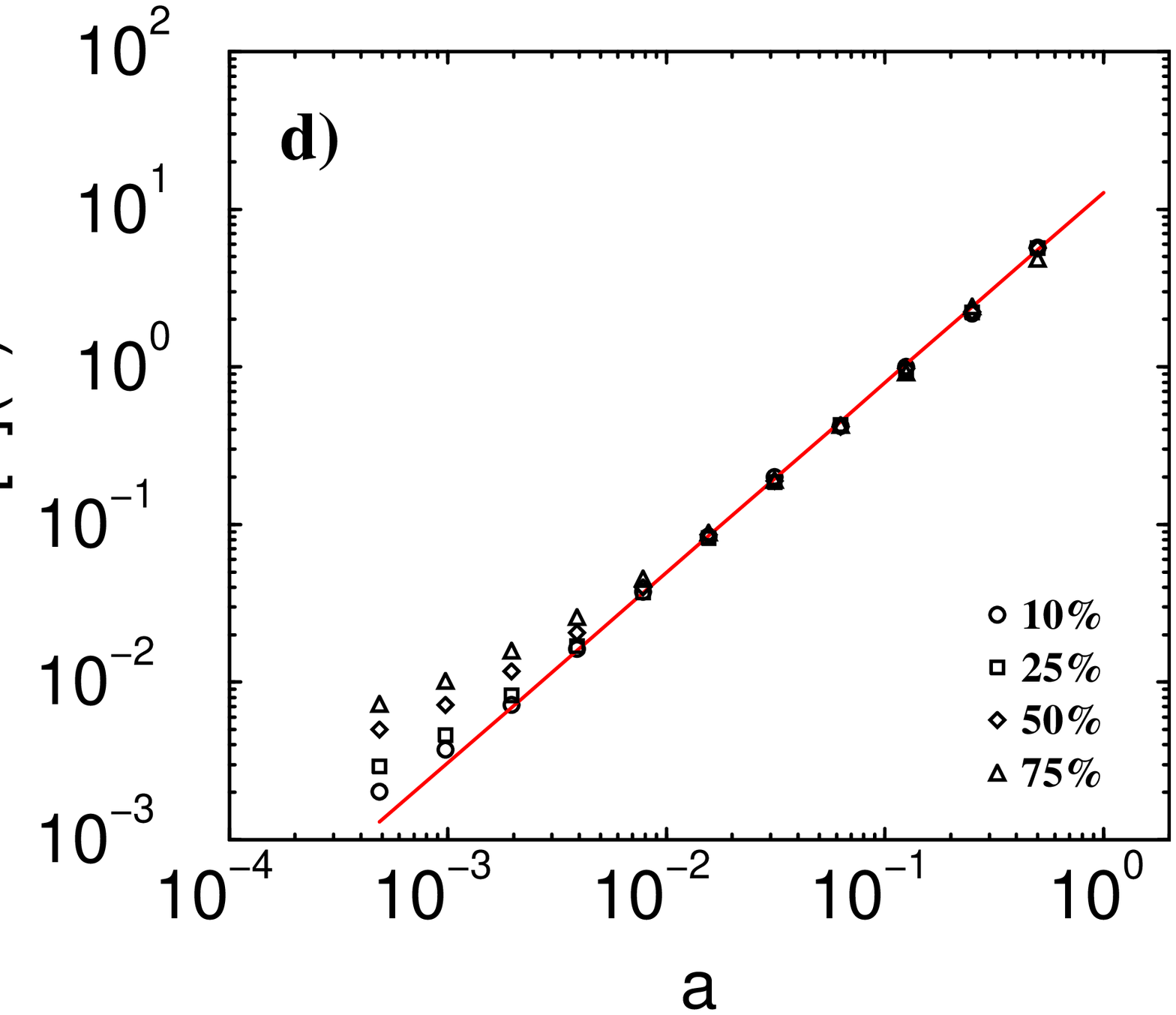,width=8.5cm,height=8.5cm} 
        \end{tabular}
    \end{center}
    \mycaption{\myauthor}{\mytitle}
\end{figure}

\begin{figure}
    \begin{center}
        \begin{tabular}{@{}c@{\hspace{1.0cm}}c@{}}
            \epsfig{file=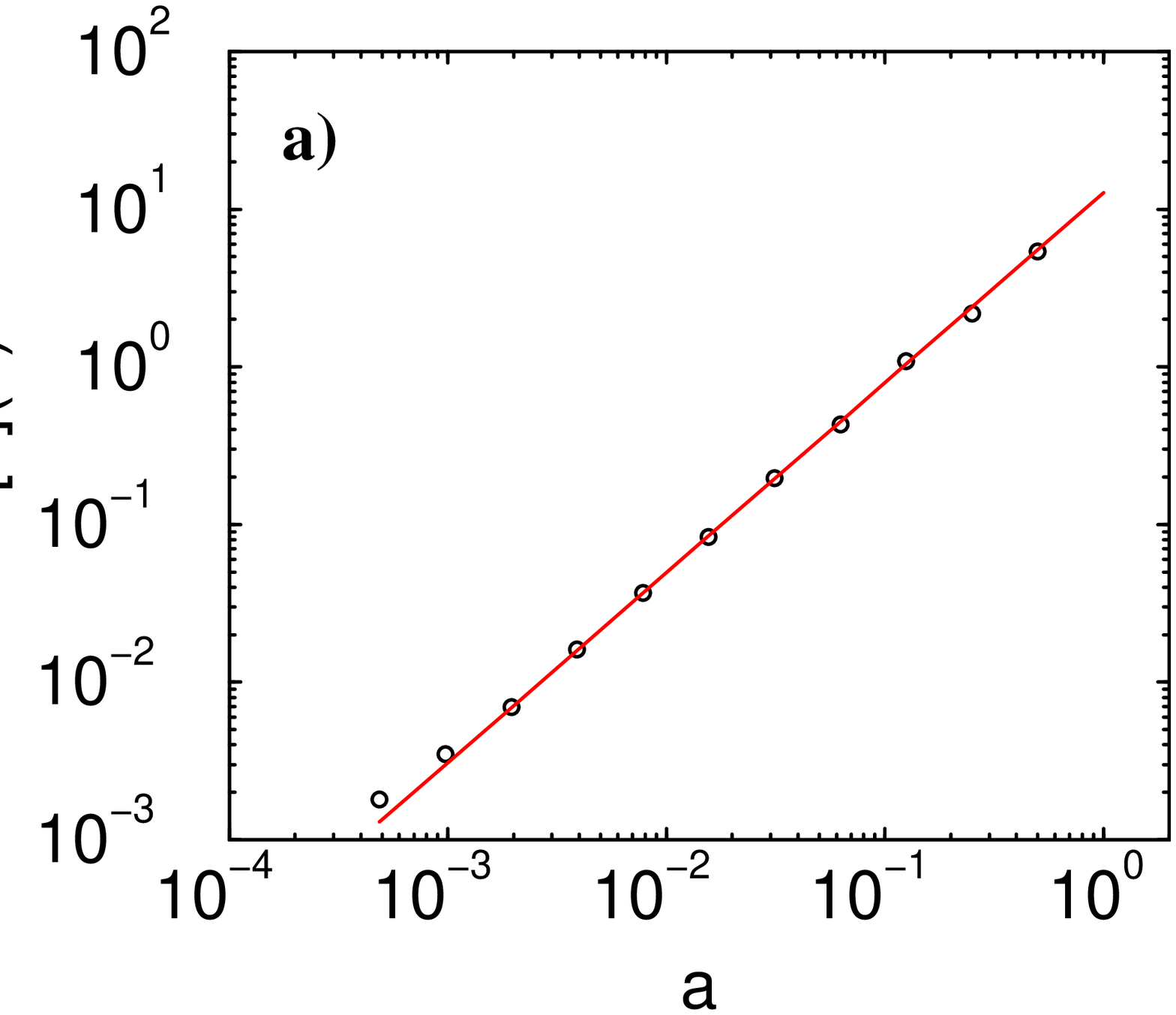,width=8.5cm,height=8.5cm} &
            \epsfig{file=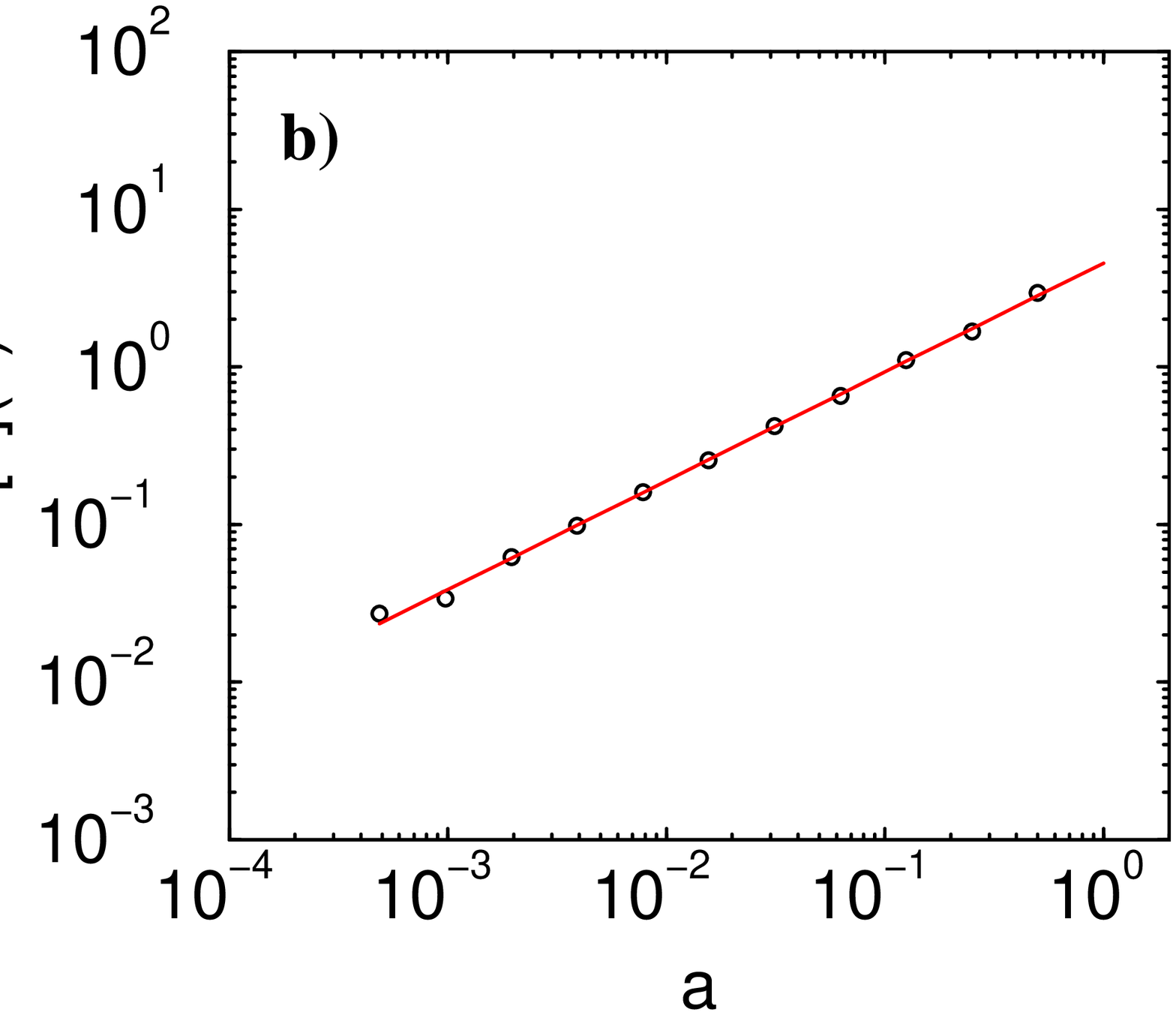,width=8.5cm,height=8.5cm} \\
            \epsfig{file=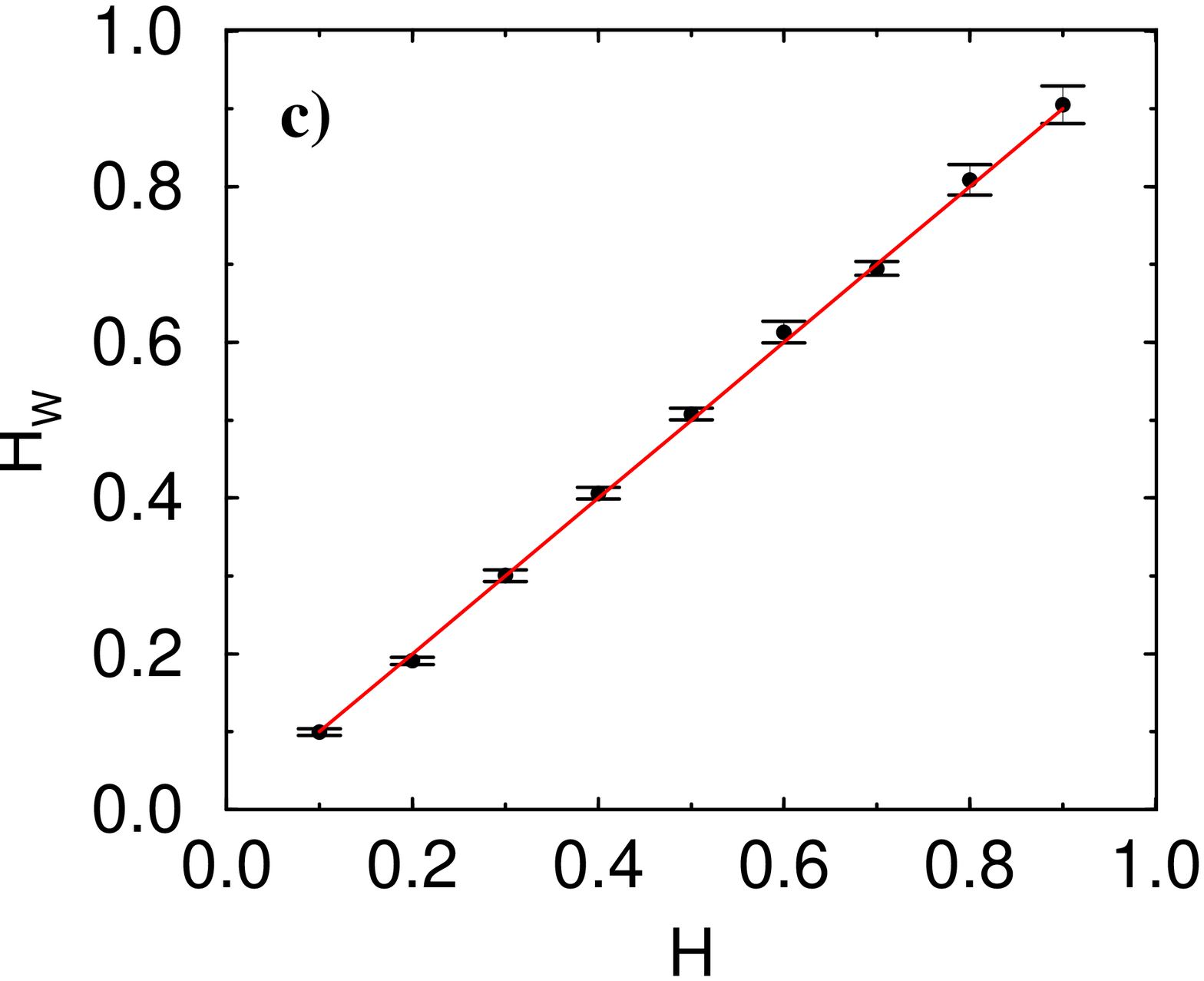,width=8.5cm,height=8.5cm} &
            \epsfig{file=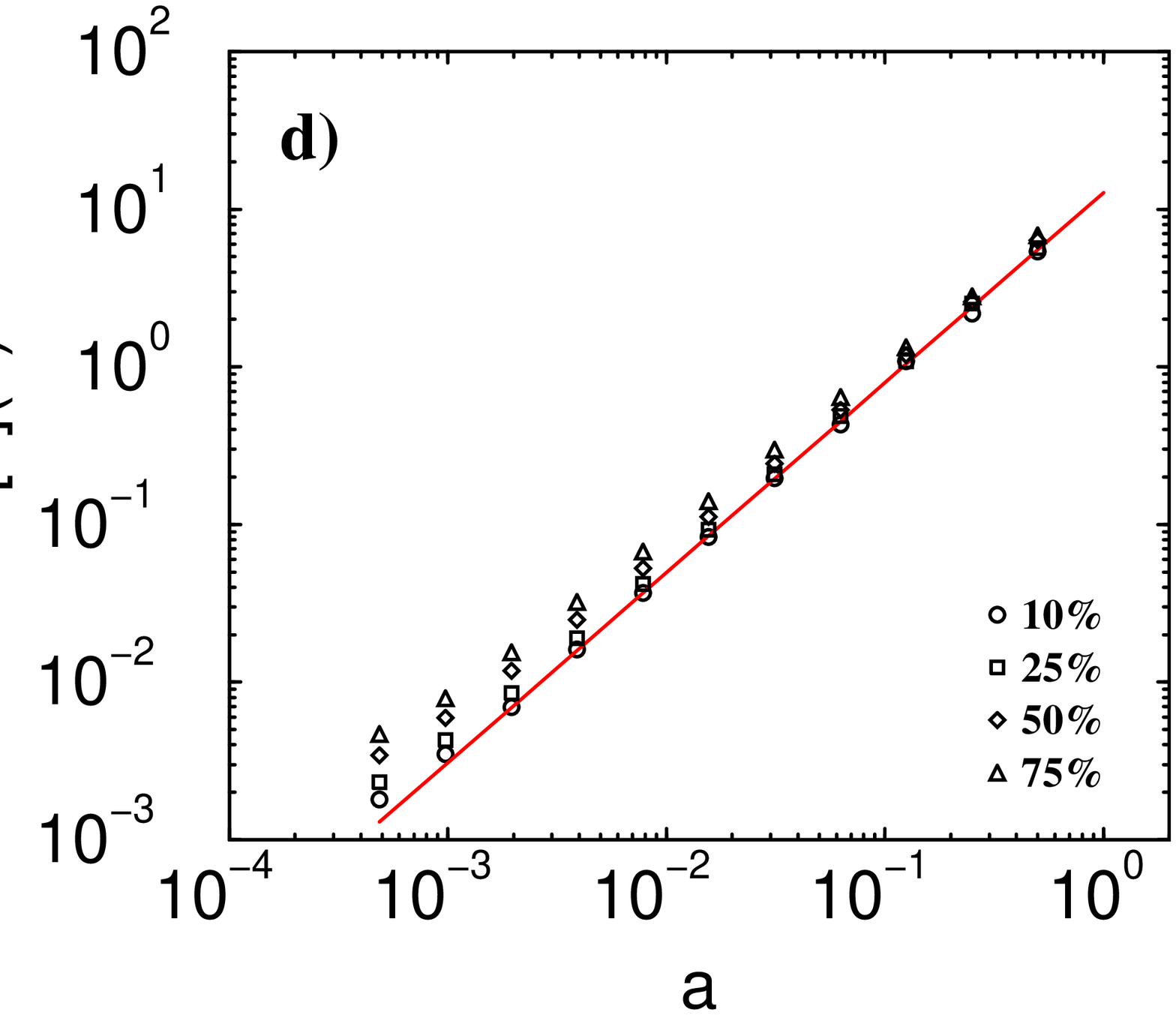,width=8.5cm,height=8.5cm} 
        \end{tabular}
    \end{center}
    \mycaption{\myauthor}{\mytitle}
\end{figure}

\begin{figure}
    \begin{center}
        \begin{tabular}{@{}c@{\hspace{1.0cm}}c@{}}
        \epsfig{file=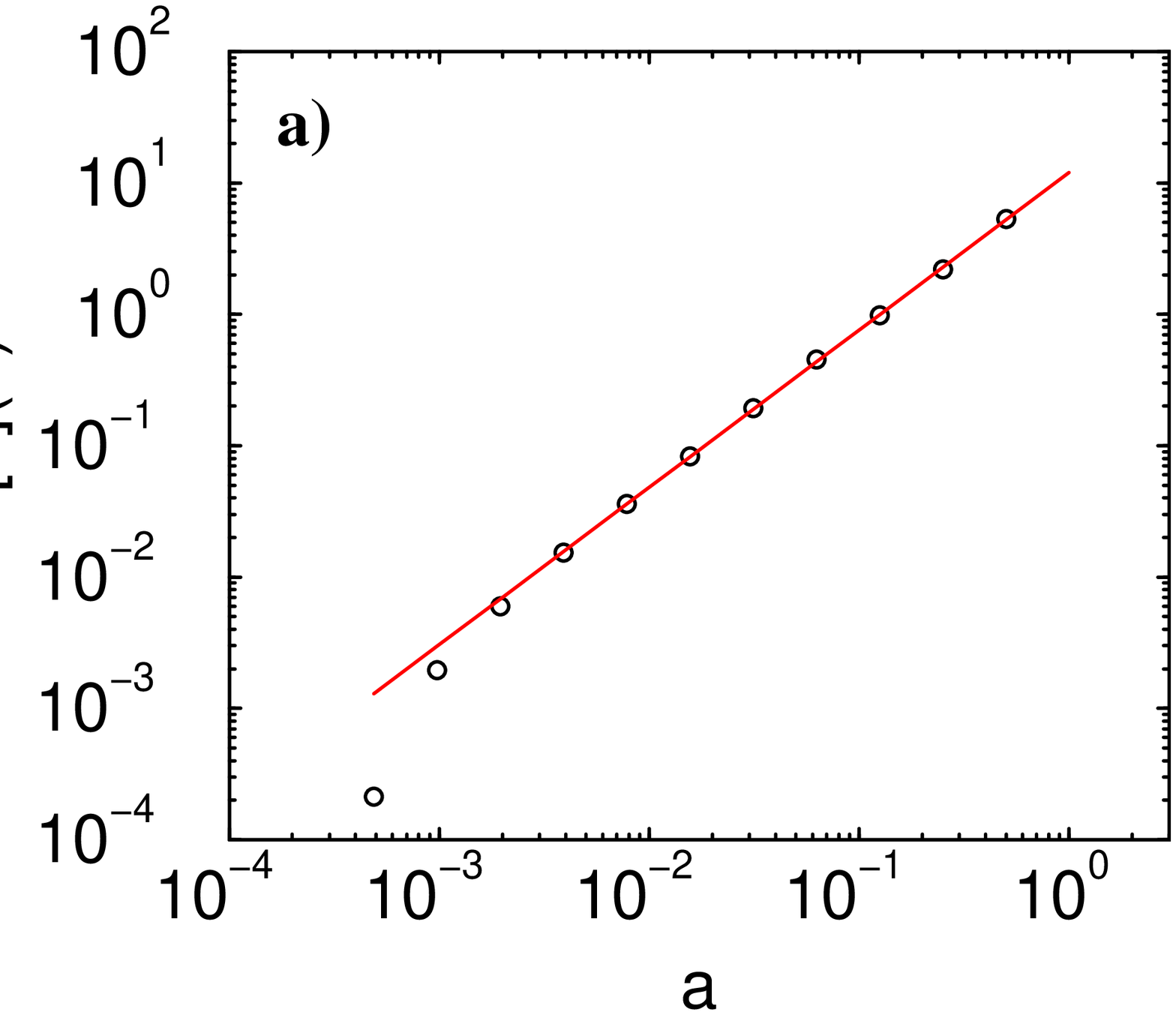,width=8.5cm,height=8.5cm} &
        \epsfig{file=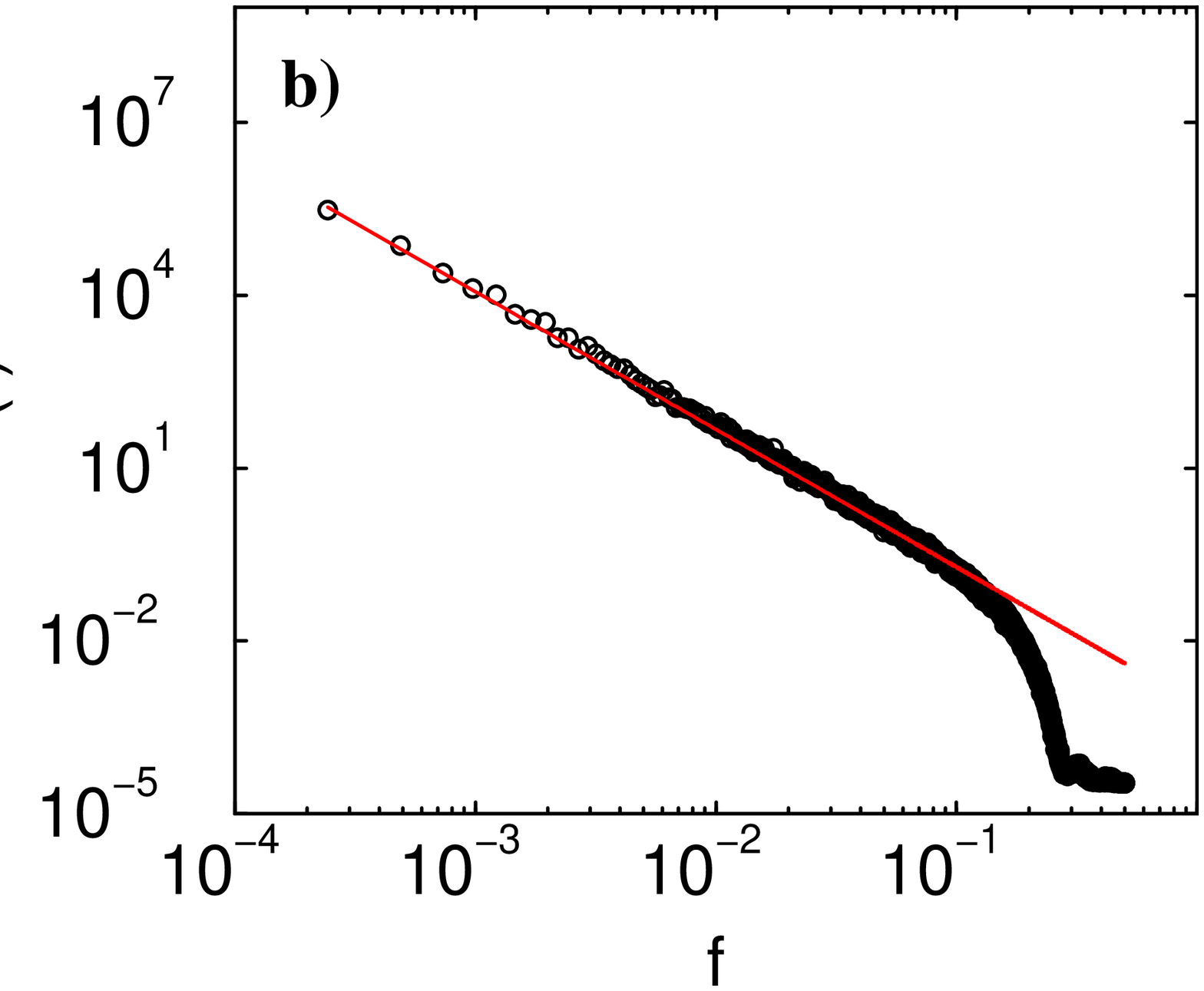,width=8.5cm,height=8.5cm} \\
        \epsfig{file=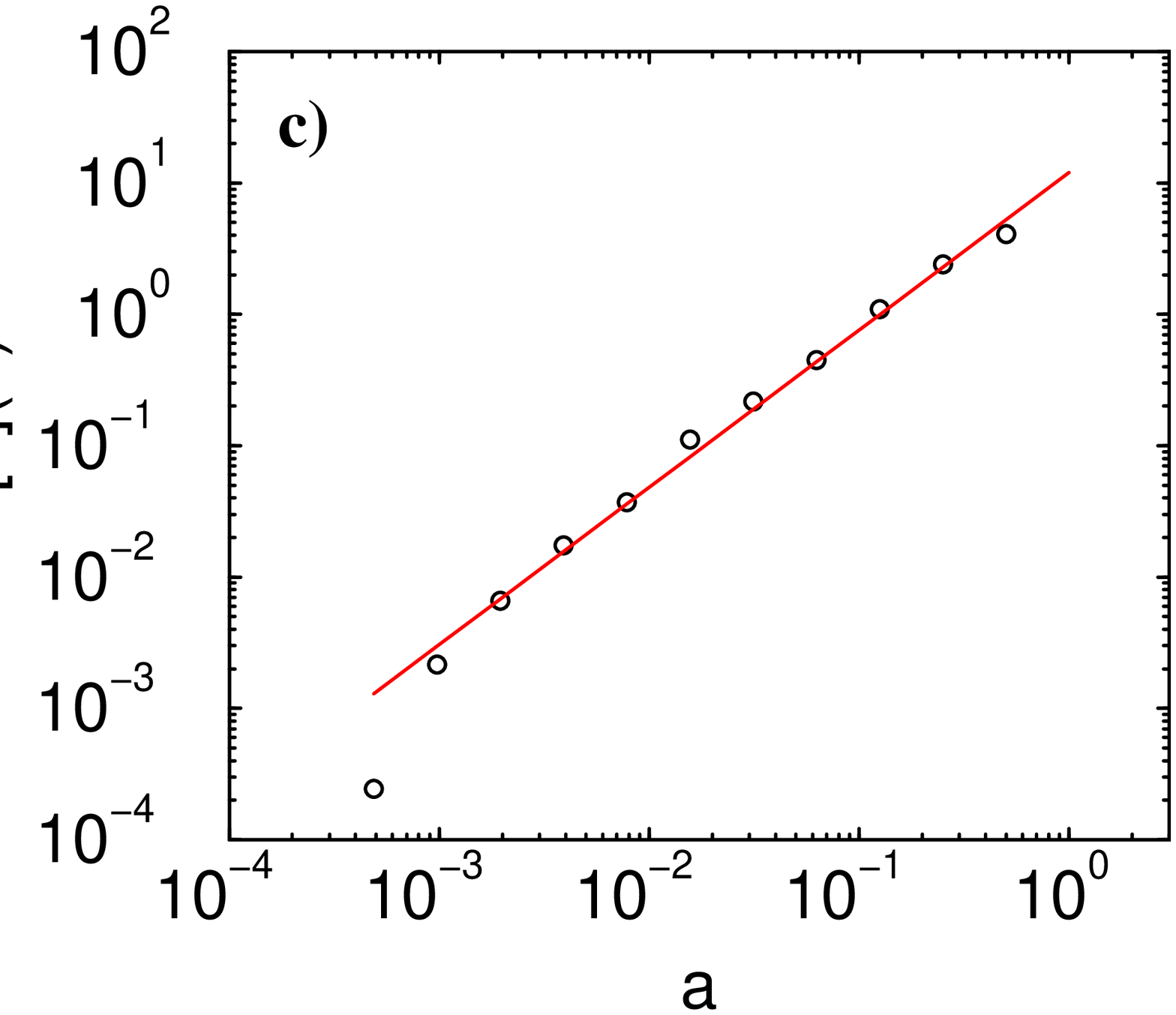,width=8.5cm,height=8.5cm}   &
        \epsfig{file=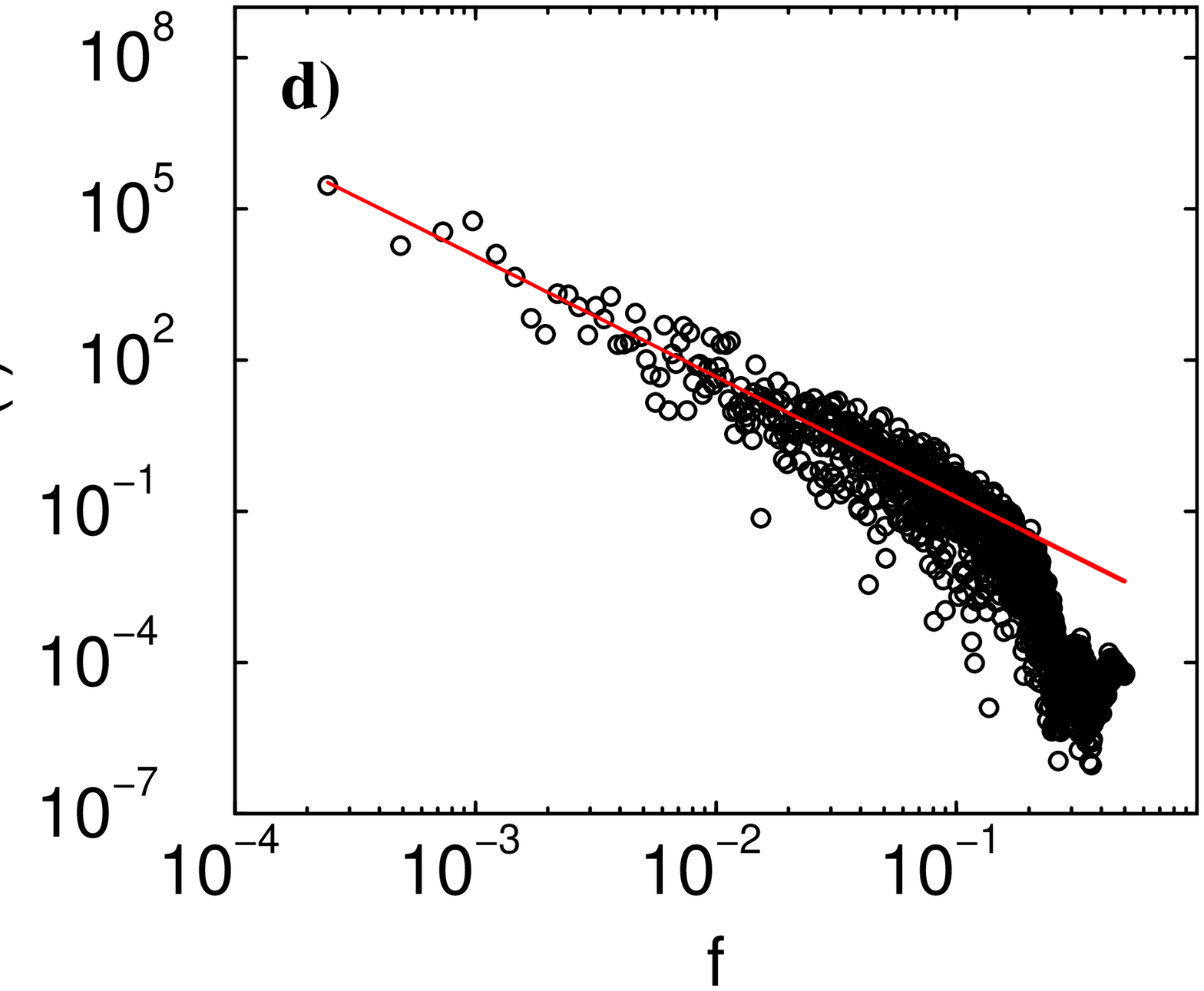,width=8.5cm,height=8.5cm} 
        \end{tabular}
    \end{center}
    \mycaption{\myauthor}{\mytitle}
\end{figure}

\begin{figure}
    \begin{center}
        \begin{tabular}{@{}c@{\hspace{1.0cm}}c@{}}
         \epsfig{file=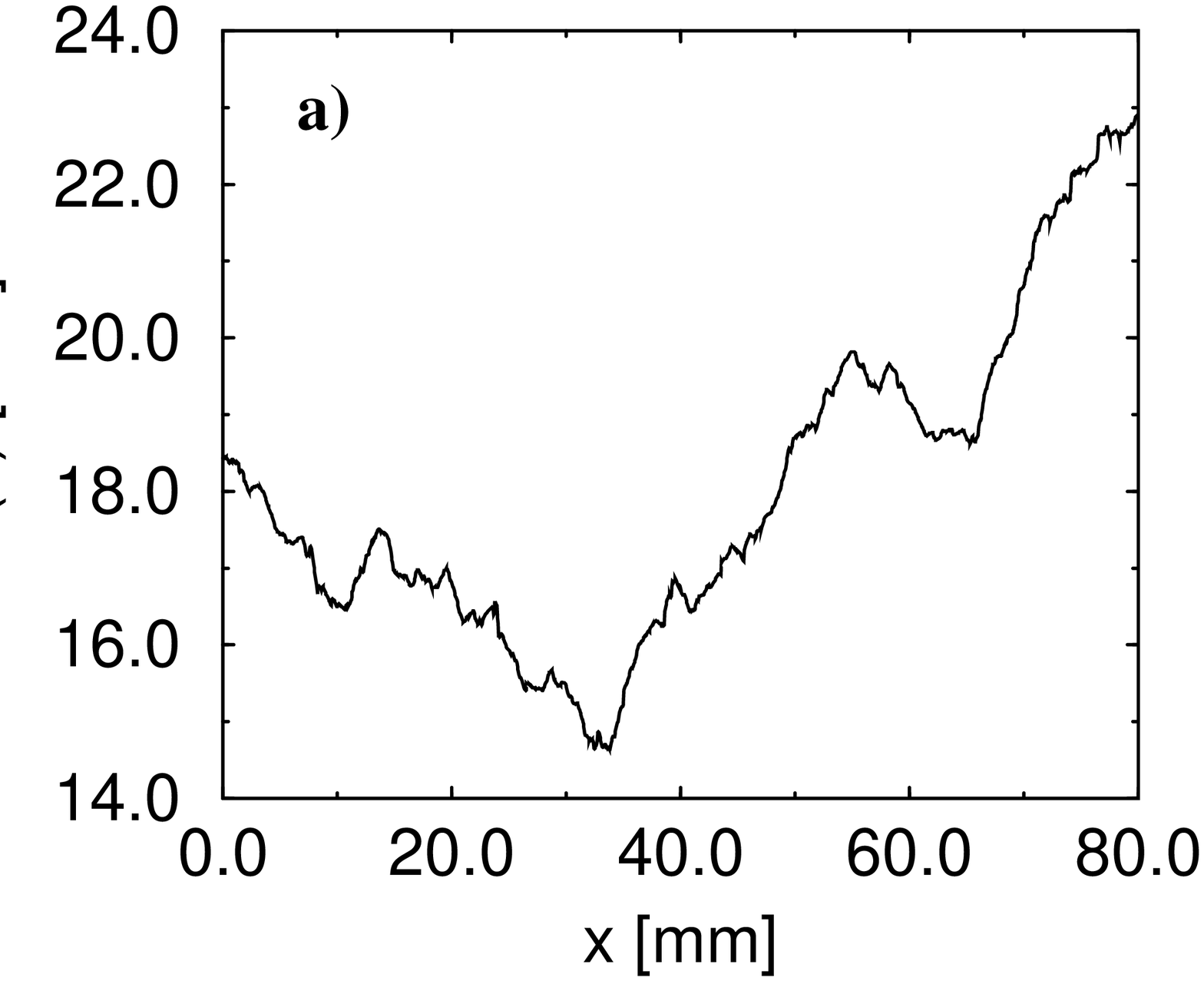,width=8.5cm,height=8.5cm} &
         \epsfig{file=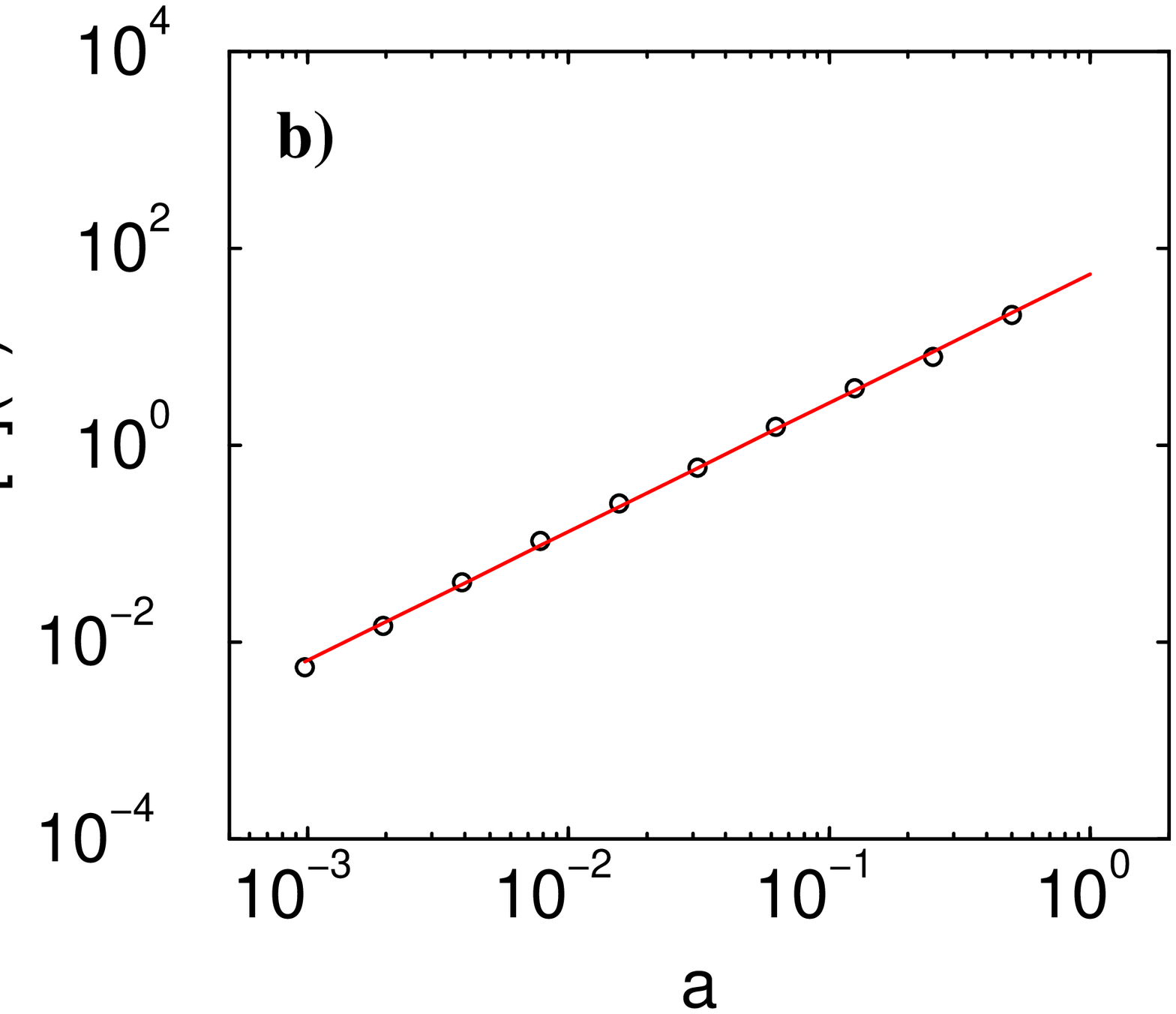,width=8.5cm,height=8.5cm} \\
         \epsfig{file=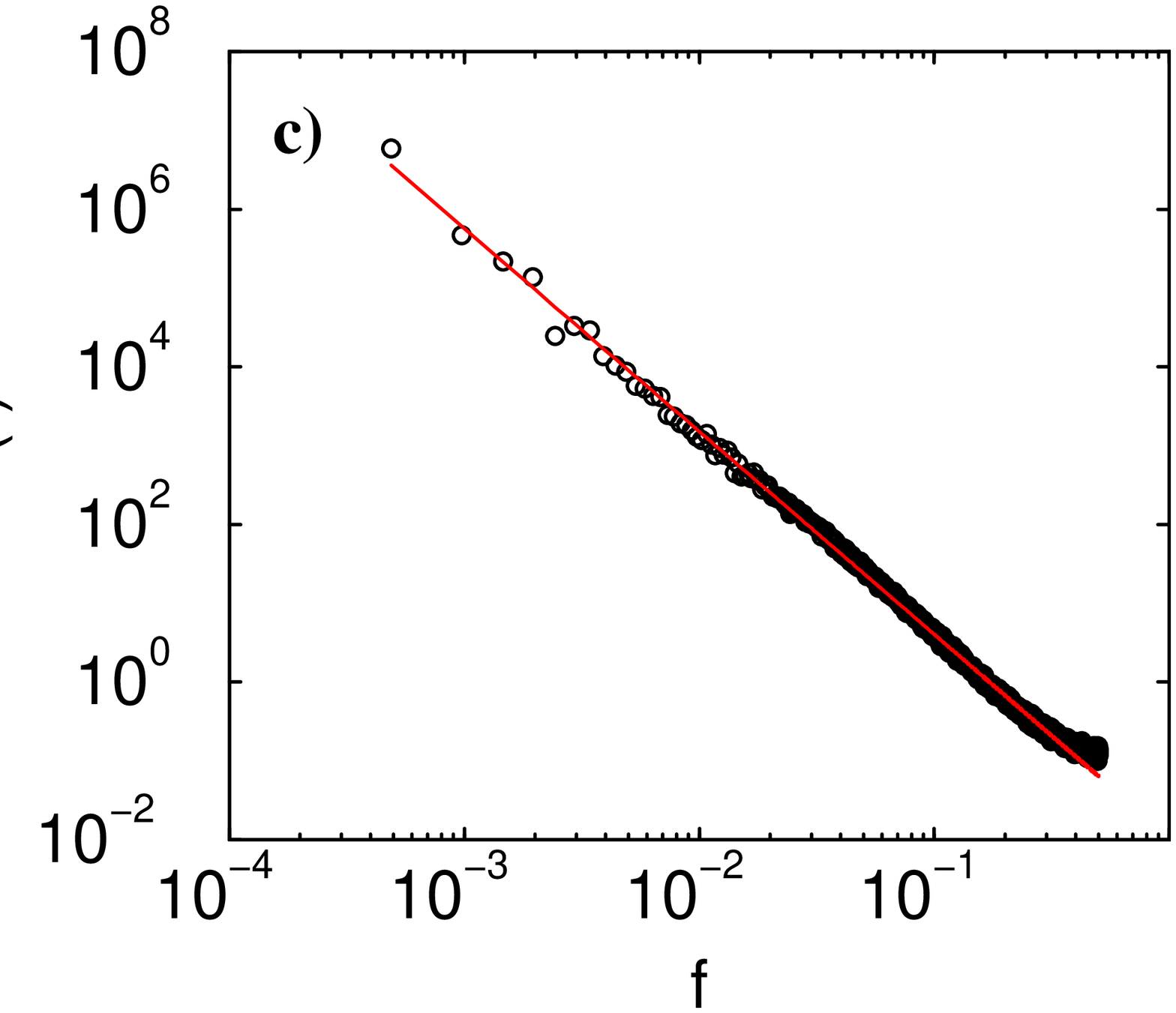,width=8.5cm,height=8.5cm} &
        \end{tabular}
    \end{center}
    \mycaption{\myauthor}{\mytitle}
\end{figure}

\begin{figure}
    \begin{center}
        \begin{tabular}{@{}c@{\hspace{1.0cm}}c@{}}
            \epsfig{file=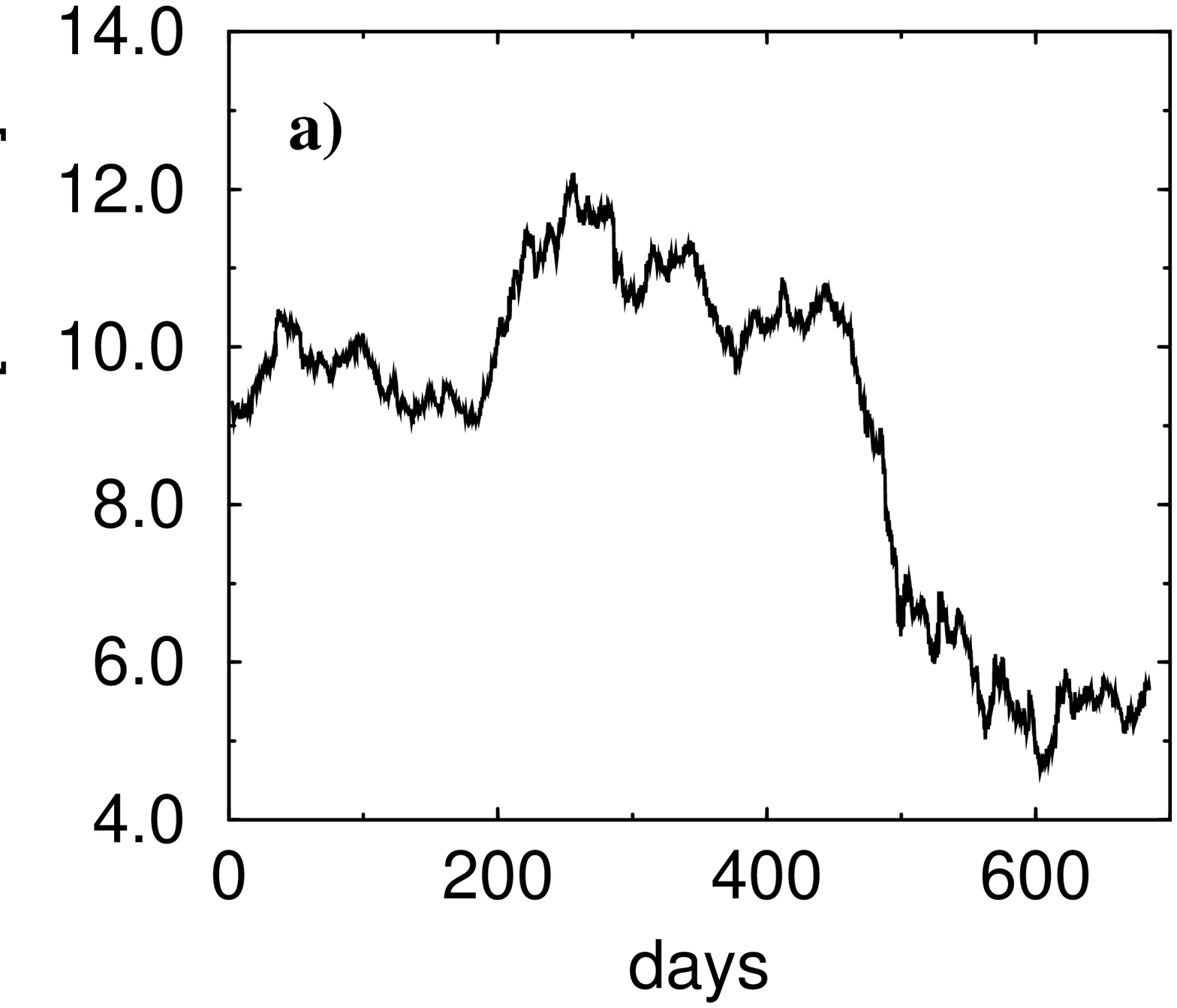,width=8.5cm,height=8.5cm} \\
            \epsfig{file=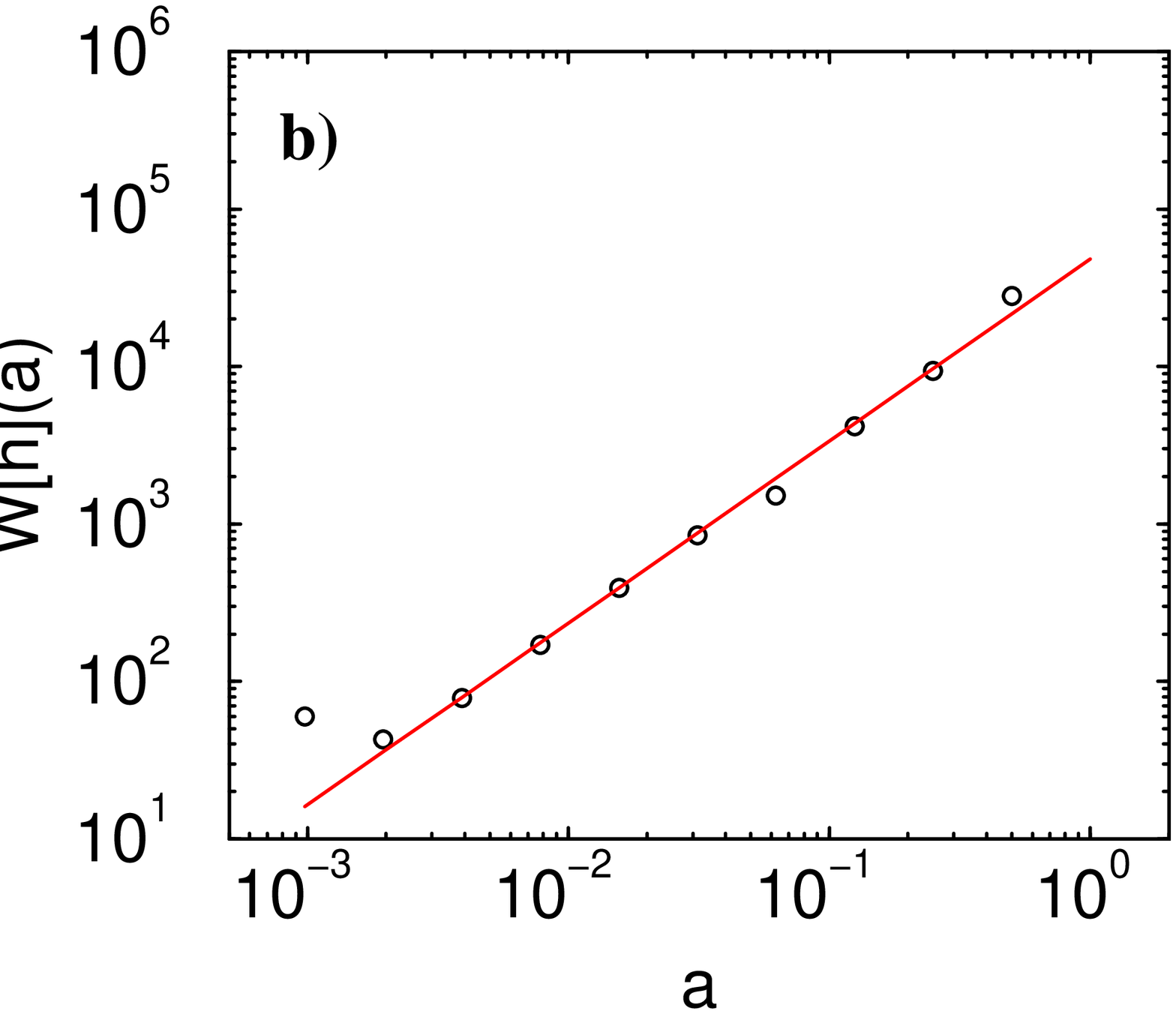,width=8.5cm,height=8.5cm}  &
            \epsfig{file=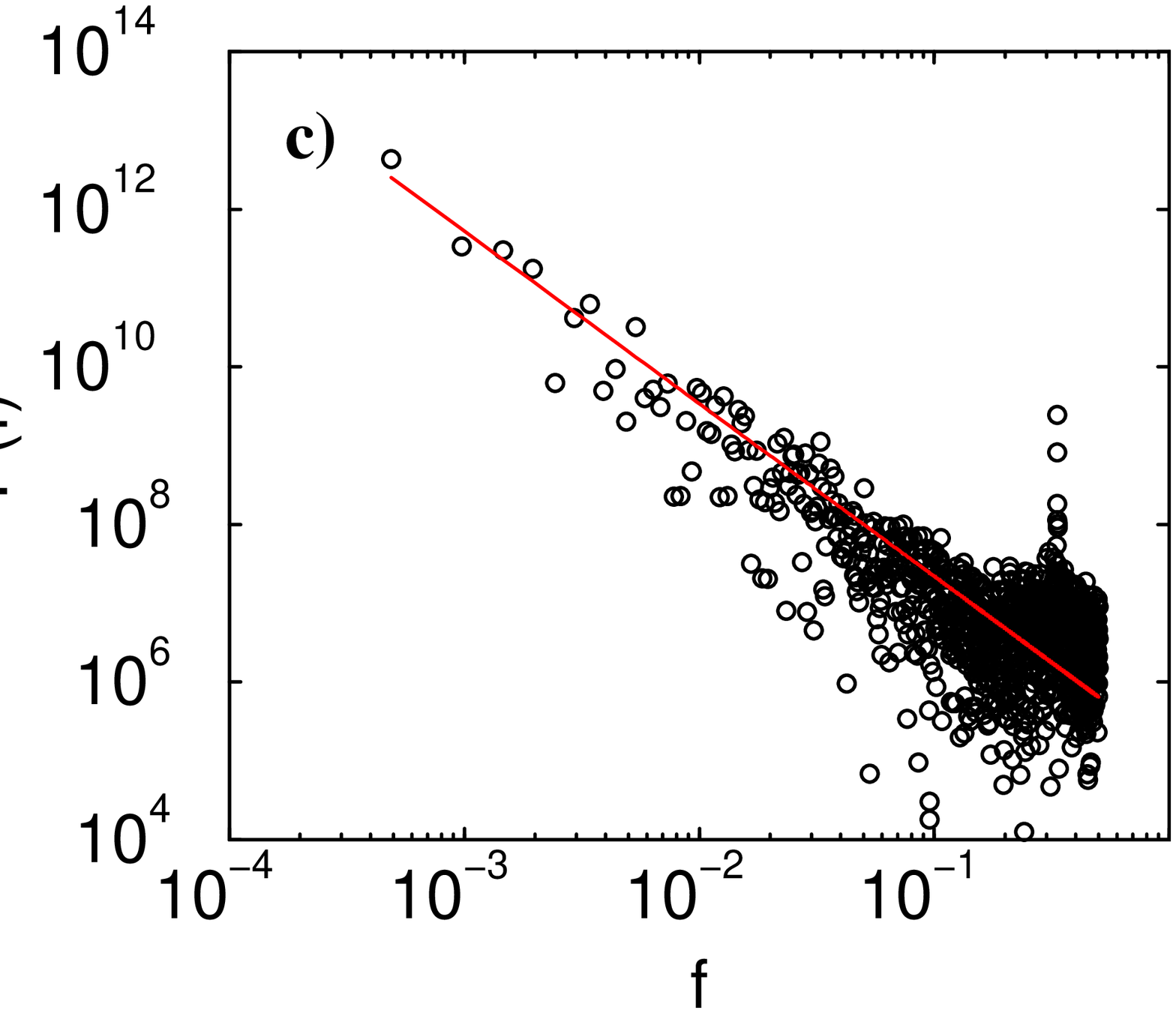,width=8.5cm,height=8.5cm}
        \end{tabular}
    \end{center}
    \mycaption{\myauthor}{\mytitle}
\end{figure}

\end{document}